\documentclass[twocolumn]{aastex63}

\usepackage{natbib}
\usepackage{color}
\usepackage{mathtools}
\usepackage{hyperref} 

\def    \apjl  		{\rm {ApJL}}

\def    \apjl  		{\rm {ApJL}}

\def	\cm		{\,{\rm {cm}}}
\def	\K		{\,{\rm K}}
\def	\g		{\,{\rm {g}}}
\def	\mum	{\,{\mu \rm{m}}}

\def \bea {\begin{eqnarray}}
\def \ena {\end{eqnarray}}

\def	\B	{{\rm B}}

\def	\C	{{\rm C}}

\def	\cm	{\,{\rm cm}}
\def	\d	{{\rm d}}

\def	\D	{{\rm D}}

\def	\erg	{\,{\rm erg}}

\def	\g	{\,{\rm g}}
\def	\gas	{\,{\rm gas}}
\def	\G	{{\rm G}}

\def	\H	{{\rm H}}

\def	\N	{{\rm N}}

\def	\s	{\,{\rm s}}

\def	\AU	{{\rm AU}}

\def	\V	{{\rm V}}

\def	\rad	{{\rm rad}}

\def    \gas     	{{\rm gas}}

\begin{document}
\shorttitle{Grain alignment and disruption in dense clouds}
\shortauthors{Hoang et al.}
\title{Grain alignment and disruption by radiative torques in dense molecular clouds and implication for polarization holes}

\author{Thiem Hoang}
\affil{Korea Astronomy and Space Science Institute, Daejeon 34055, Republic of Korea, \href{mailto:thiemhoang@kasi.re.kr}{thiemhoang@kasi.re.kr}}
\affil{Korea University of Science and Technology, 217 Gajeong-ro, Yuseong-gu, Daejeon, 34113, Republic of Korea}

\author{Le Ngoc Tram}
\affil{Stratospheric Observatory for Infrared Astronomy, Universities Space Research Association, NASA Ames Research Center, MS 232-11, Moffett Field, 94035 CA, USA}

\author{Hyeseung Lee}
\affil{Korea Astronomy and Space Science Institute, Daejeon 34055, Republic of Korea}

\author{Pham Ngoc Diep}
\affil{Vietnam National Space Center, Vietnam Academy of Science and Technology, 18 Hoang Quoc Viet, Hanoi, Vietnam}

\author{Nguyen Bich Ngoc}
\affil{Vietnam National Space Center, Vietnam Academy of Science and Technology, 18 Hoang Quoc Viet, Hanoi, Vietnam}

\begin{abstract}
Dust polarization induced by aligned grains is widely used to study magnetic fields in various environments, including star-forming regions. However, the question of to what optical depth grain alignment still exists in a dense molecular cloud (MC) is unclear. In this paper, we aim to achieve analytical formulae for the minimum size of aligned grains ($a_{\rm align}$) and rotational disruption ($a_{\rm disr}$) by RAdiative Torques (RATs) as a function of the local physical parameters within dense MCs. We first find the analytical approximations for the radiation strength and the mean wavelength of the attenuated radiation field in a dense MC without and with embedded stars and then derive analytical formulae for $a_{\rm align}$ and $a_{\rm disr}$ as functions of the visual extinction $A_{V}$ and the gas density. We find that within a starless core of density $n_{\rm H}\sim 10^{5}\cm^{-3}$, grains of size $a<0.25\mum$ can be aligned up to $A_{V}\sim 5$ by RATs, whereas micron-sized grains can still be aligned at $A_{V}\sim 50$. The increase in the alignment size with $A_{V}$ can explain the presence of polarization holes observed toward starless cores. For MCs with an embedded protostar, we find that the efficiency of both alignment and rotational disruption increases toward the protostar due to the increasing radiation strength. Such a disruption effect results in the decrease of the polarization degree with $A_{V}$ or emission intensity, which reproduces the popular polarization holes observed toward the location of protostars. Finally, we derive the formula for the maximum $A_{V}$ where grain alignment still exists in a starless core and discuss its potential for constraining grain growth.

\end{abstract}

\section{Introduction}\label{sec:intro}
Interstellar dust is an essential component of the interstellar medium (ISM) and plays an important role in various astrophysical processes, including gas heating, star and planet formation, and grain-surface chemistry (see \citealt{2003ARA&A..41..241D} for a review). Dust polarization induced by grain alignment allows us to study magnetic fields in various astrophysical environments, from the diffuse interstellar medium (ISM) to dense molecular clouds (MCs) to protoplanetary disks. Magnetic fields are thought to play an important role in the process of star formation \citep{2012ARA&A..50...29C}. Therefore, observing dust polarization in star-forming regions is valuable to clarify the role of magnetic fields in the process (\citealt{2019FrASS...6...15P}). However, the question is to what optical depth in a dense MC grain alignment still exists, enabling a robust detection of the magnetic fields.

Observations of background starlight polarization in optical-near infrared (NIR) are usually used to probe grain alignment in dark MCs (e.g., \citealt{2008ApJ...674..304W}). Optical-NIR observations toward starless cores usually reveal the decrease of the polarization fraction with the visual extinction (\citealt{1992ApJ...399..108G}; \citealt{2008ApJ...674..304W}), which is known as "polarization hole". Polarization hole is also observed in far-IR/submm toward prestellar cores (\citealt{WardThompson:2000p6330}; \citealt{Crutcher:2004p6727}). The exact origin of the polarization hole is still debated, but the popular explanation is the decrease of grain alignment toward the central region due to high gas density (\citealt{{1992ApJ...399..108G},{1995ApJ...448..748G}}). Some observations of dust polarization from starless cores indeed reveal the loss of grain alignment at a large optical depth of $A_{V}>20$ (\citealt{2014A&A...569L...1A}; \citealt{2015AJ....149...31J}, whereas other observations (\citealt{2017ApJ...849..157W}; \citealt{2019ApJ...880...27P}) show that grain alignment still exists at such large $A_{V}$. 

Moreover, NIR observations of background starlight polarization toward starless dark clouds reveal unusually large values of the peak wavelength (i.e., the wavelength of the maximum polarization), $\lambda_{\rm max}\sim 1-1.2\mum$ (\citealt{2016ApJ...833..176C}), and $\lambda_{\max}\sim 0.6-0.9\mum$ (\citealt{2017ApJ...849..157W}). This suggests that the grain sizes are much large than in the diffuse ISM, implying significant grain growth in dense MCs. Observations of polarized dust emission in far-IR/submm also reveal the existence of grain alignment in starless cores at large extinction of $A_{V}\gtrsim 20$. Moreover, observations of scattered light in NIR from dense clouds irradiated by a nearby star show evidence of micron-sized grains, which is known as core shines (\citealt{2010Sci...329.1622P}; \citealt{2010A&A...511A...9S}; \citealt{2012A&A...541A..12J}; \citealt{Ysard:2013fn}). Theoretically, grain growth is expected to occur in dense MCs due to the accretion of gas species to the grain surface and grain-grain collisions \citep{2013MNRAS.434L..70H}. The remaining question is how such dust grains can still be aligned at such large values of $A_{V}$. 

The problem of grain alignment is one of the most long standing questions in astrophysics. Since the discovery of starlight polarization in 1949 (\citealt{Hall:1949p5890}; \citealt{Hiltner:1949p5856}) that revealed the alignment of interstellar grains, many theories have been proposed (see \citealt{2003JQSRT..79..881L} for a review). The popular theory of grain alignment is Radiative Torque (RAT) theory (\citealt{1976Ap&SS..43..291D}; \citealt{1997ApJ...480..633D}; \citealt{2007MNRAS.378..910L}; \citealt{Hoang:2008gb}), which is supported by numerous observations (\citealt{2015ARA&A..53..501A} and \citealt{LAH15} for recent reviews). 
Numerical modeling of grain alignment by RATs in a starless dark cloud was presented in \cite{2005ApJ...631..361C} and \cite{2007ApJ...663.1055B}. Both studies only consider the alignment of grains induced by attenuated ISRF and assume a maximum surface-to-center optical depth of $A_{V}=10$. Therefore, their results cannot be directly applied to interpret observational data at larger $A_{V}$ in the era of high spatial resolution observations. The first goal of this paper is to derive an analytical formula for the minimum size of aligned grains for dense MCs having much larger $A_{V}$.

Protostars are thought to form in the dense core of MCs. Polarimetric observations toward protostars by single dish and interferometric observations (e.g., \citealt{2013ApJ...771...71L}; \citealt{2014ApJS..213...13H}; \citealt{2018ApJ...855...92C}; see \citealt{2019FrASS...6...15P}) usually reveal the existence of "polarization hole", which is described as the decrease of the polarization fraction with the column density or intensity of dust emission. Previous studies usually appeal to two possible reasons to explain the polarization hole, including the tangling of the magnetic field and the decrease of alignment efficiency toward the protostar (\citealt{2014ApJS..213...13H}). However, in the RAT alignment framework, one expects the increase of grain alignment efficiency toward the protostar due to the increasing incident radiation flux, resulting in the increase (instead of the decrease) of the polarization fraction with the peak intensity. Therefore, the underlying origin of the polarization hole toward protostars is difficult to reconcile in terms of grain alignment theory. 

Recently, a new mechanism of grain destruction based on centrifugal force, so-called Radiative Torque Disruption (RATD) is introduced (\citealt{Hoang:2019da}; \citealt{2019ApJ...876...13H}, which is particularly efficient near protostars (see \citealt{Hoang:2020} for a review). The RATD mechanism disrupts large grains into smaller ones, resulting in the decrease of the polarization fraction with increasing the incident radiation field (\citealt{2020ApJ...896...44L}). As a result, we expect that the joint action of grain alignment and disruption by RATs could explain the observed polarization toward protostars. The second goal of this paper is to understand the origin of polarization holes by detailed modeling of grain alignment and rotational disruption by RATs in a dense MC with a central protostar.

The structure of the paper is as follows. In Section \ref{sec:RATs}, we describe RATs and derive analytical formulae for the average RATs over a radiation spectrum. In Section \ref{sec:align_disrupt}, we present analytical formulae for grain alignment and disruption as functions of the visual extinction and local physical parameters. In Sections \ref{sec:GMC} and \ref{sec:protostar}, we apply our analytical formulae for a molecular cloud without and with an embedded source. Discussion of our results for polarization holes and grain growth is presented in Section \ref{sec:discuss}. Our main findings are summarized in Section \ref{sec:summary}.

\section{Radiative torques of irregular grains}\label{sec:RATs}
In this section, we describe RATs of irregular grains and derive analytical formulae for the RAT efficiency averaged over an arbitrary radiation spectrum, which will be used for alignment and disruption studies.
\subsection{Radiative torques}
Let $u_{\lambda}$ be the spectral energy density and $\gamma$ be the anisotropy degree of the radiation field.
The energy density of the radiation field is then 
\bea
u_{\rad}=\int_{0}^{\infty} u_{\lambda}d\lambda.\label{eq:urad}
\ena

To describe the strength of a radiation field, we introduce the dimensionless parameter $U=u_{\rm rad}/u_{\rm ISRF}$ with 
$u_{\rm ISRF}=8.64\times 10^{-13}\erg\cm^{-3}$ being the energy density of the average interstellar radiation field (ISRF) in the solar neighborhood as given by \cite{1983A&A...128..212M} (hereafter MMP83). Thus, the typical value for the ISRF in the solar neighborhood is $U=1$. 

Let $a$ be the effective size of the grain which is defined as the radius of the sphere with the same volume as the irregular grain. Radiative torque (RAT) induced by the interaction of an anisotropic radiation field with the irregular grain is defined as
\bea
{\Gamma}_{\lambda}=\pi a^{2}
\gamma u_{\lambda} \left(\frac{\lambda}{2\pi}\right){Q}_{\Gamma},\label{eq:GammaRAT}
\ena
where ${Q}_{\Gamma}$ is the RAT efficiency (\citealt{1996ApJ...470..551D}; \citealt{2007MNRAS.378..910L}).

The magnitude of RAT efficiency can be approximated by a power-law
\bea
Q_{\Gamma}= \alpha\left(\frac{{\lambda}}{a}\right)^{-\eta},\label{eq:QAMO}
\ena
where $\alpha$ and $\eta$ are the constants that depend on the grain size, shape, and optical constants. Numerical calculations of RATs for several shapes of different optical constants in \cite{2007MNRAS.378..910L} find the slight difference in RATs among the realization. They adopted the coefficients $\alpha=0.4,\eta=0$ for $a_{\rm trans}<a<\lambda/0.1$, and $\alpha=2.33,\eta=3$ for $a<a_{\rm trans}$ where $a_{\rm trans}=\lambda/1.8$ denotes the transition size at which the RAT efficiency changes the slope. Thus, the maximum RAT efficiency is $Q_{\Gamma,\max}=\alpha$. 

An extensive study for a large number of irregular shapes by \cite{2019ApJ...878...96H} shows moderate difference in RATs for silicate, carbonaceous, and iron compositions. Moreover, the analytical formula (Equation \ref{eq:QAMO}) is also in good agreement with their numerical calculations. Therefore, one can use Equation (\ref{eq:QAMO}) for the different grain compositions and grain shapes, and the difference is of order unity.

The radiative torque averaged over the incident radiation spectrum of spectral energy density $u_{\lambda}$ is defined as
\bea
\overline{\Gamma}_{\rm RAT}&=&\pi a^{2}
\gamma u_{\rad} \left(\frac{\overline{\lambda}}{2\pi}\right)\overline{Q}_{\Gamma},\label{eq:GammaRAT_num}
\ena
where
\bea
\overline{Q}_{\Gamma} = \frac{\int_{0}^{\infty} \lambda Q_{\Gamma}u_{\lambda} d\lambda}{\int_{0}^{\infty} \lambda u_{\lambda} d\lambda},~\bar{\lambda}=\frac{\int_{0}^{\infty} \lambda u_{\lambda}d\lambda}{u_{\rm rad}}\label{eq:Qavg_num}
\ena
are the average RAT efficiency and the mean wavelength, respectively.

In general, one can numerically calculate the average RAT and average RAT efficiency using Equations (\ref{eq:GammaRAT_num}) and (\ref{eq:Qavg_num}) for an arbitrary radiation spectrum $u_{\lambda}$. To facilitate analysis of grain alignment from observations, in the following, we will derive analytical formulae for $\overline{Q}_{\Gamma}$ and $\bar{\lambda}$ for two popular radiation fields.

\subsection{Average RAT over a stellar radiation spectrum}

For a radiation field produced by a star of temperature $T_{\star}$, the spectral energy density at distance $d$ from the star is
\bea
u_{\lambda}(T_{\star})=\frac{4\pi R_{\star}^{2}F_{\lambda}}{4\pi d^{2}c}=\frac{\pi B_{\lambda}(T_{\star})}{c}\left(\frac{R_{\star}}{d}\right)^{2},
\ena
where $F_{\lambda}=\pi B_{\lambda}$ is the radiation flux from the stellar surface of radius $R_{\star}$, and $B_{\lambda}=(2hc^{2}/\lambda^{5}) (\exp(hc/\lambda kT_{\star})-1)^{-1}$ is the Planck function. For simplicity, we first disregard the reddening effect by intervening dust.

The mean wavelength of the stellar radiation field is given by
\bea
\bar{\lambda}(T_{\star})&&=\frac{\int_{0}^{\infty} \lambda B_{\lambda}(T_{\star})d\lambda}{\int B_{\lambda}(T_{\star})d\lambda}=
\left(\frac{2\pi k^{3}\zeta(3)\Gamma(3)}{\sigma ch^{2}}\right)\frac{1}{T_{\star}}\nonumber\\
&&\simeq \frac{0.53\cm \K}{T_{\star}},\label{eq:wavemean_star}
\ena
where $\zeta$ and $\Gamma$ are the Riemann and Gamma functions.
 
The radiation energy density of the stellar radiation field becomes
\bea
u_{\rm rad}(T_{\star})=\left(\frac{R_{\star}}{d}\right)^{2}\frac{\int_{0}^{\infty} \pi B_{\lambda}(T_{\star})d\lambda}{c}=\left(\frac{R_{\star}}{d}\right)^{2}\frac{\sigma T_{\star}^{4}}{c}.
\ena

For small grains of $a<\bar{\lambda}/1.8$, plugging $Q_{\Gamma}$ from Equation (\ref{eq:QAMO}) and $u_{\lambda}(T_{\star})$ into Equation (\ref{eq:Qavg_num}), one obtains the following after taking the integral,
\bea
\overline{Q}_{\Gamma}&=&\frac{2\pi \alpha k^{\eta+3}}{\sigma h^{\eta+2}c^{n+1}}\left(\frac{\zeta(3)\Gamma(3)2\pi k^{3}}{\sigma ch^{2}} \right)^{\eta-1}\zeta(\eta+3)\Gamma(\eta+3)\nonumber\\
&&\times\left(\frac{\bar{\lambda}}{a}\right)^{-\eta}.\label{eq:Qmean_star}
\ena

Plugging the RAT parameters of $\alpha=2.33$ and $\eta=3$ into Equation (\ref{eq:Qmean_star}), one obtains
\bea
\overline{Q}_{\Gamma}\simeq 
6\left(\frac{\bar{\lambda}}{a}\right)^{-3},\label{eq:Qavg_star}
\ena 
which is applicable for $a\lesssim (Q_{\Gamma,\max}/6)^{1/3}\bar{\lambda}= \bar{\lambda}/2.5$. Large grains of $a>\bar{\lambda}/2.5$ then have $\bar{Q}_{\Gamma}=Q_{\Gamma,\max}=0.4$.


\subsection{Average RAT over the interstellar radiation field}
Following \cite{1983A&A...128..212M}, the average ISRF in the solar neighborhood can be described by three stellar radiation fields,
\bea
u_{\lambda, \rm ISRF}=\sum_{i}u_{\lambda}(T_{\star,i})=\frac{4\pi}{c}\sum_{i=1}^{3}W_{i}B_{\lambda}(T_{\star,i}),\label{eq:uwave_ISRF}
\ena
where $T_{\star,i}= 7500, 4000, 3000\K$ are the temperatures of stars, and $W_{i}=10^{-14}, 1.65\times 10^{-13}, 4\times 10^{-13}$ are the dilution factors (see \citealt{1996ApJ...470..551D}). Above, we have ignored the contribution of the enhanced ultraviolet (UV) which contributes $\sim 10\%$ of the total energy.

The RAT efficiency averaged over the radiation spectrum, $u_{\lambda,\rm ISRF}$, is then written as
\bea
\overline{\Gamma}_{\rm RAT}=\sum_{i}\Gamma_{\rm RAT}(T_{\star,i}),\label{eq:Gamma_ISRF}
\ena
where $\Gamma_{\rm RAT}(T_{\star,i})$ is the average radiative torque over the stellar radiation of star temperature $T_{\star,i}$ given by
\bea
\Gamma_{\rm RAT}(T_{\star,i})=\pi a^{2}
\gamma u_{\rad,i} \left(\frac{\overline{\lambda}_{i}}{2\pi}\right)\overline{Q}_{\Gamma,i},
\ena
where $\bar{\lambda}_{i}$ and $\overline{Q}_{\Gamma,i}$ are given by Equation (\ref{eq:wavemean_star}) and (\ref{eq:Qmean_star}), respectively, and 
\bea
u_{\rad,i}=\frac{4W_{i}}{c}\sigma T_{\star,i}^{4}.\label{eq:urad_i}
\ena

For $T_{\star,i}=7500, 4000, 3000\K$, Equation (\ref{eq:wavemean_star}) implies $\bar{\lambda}_{i}\equiv \bar{\lambda}(T_{\star,i})=0.76, 1.32,1.76\mum$, and $u_{\rad,i}=2.39\times 10^{-13},3.19\times 10^{-13},2.45\times 10^{-13}\erg\cm^{-3}$. The total radiation energy is then $u_{\rm rad,\rm ISRF}=\sum_{i}u_{\rad,i}=8.04\times 10^{-13}\erg\cm^{-3}$, which is slightly smaller than $u_{\rm ISRF}=8.64\times 10^{-13}\erg\cm^{-3}$ when accounting for the UV part.

The mean wavelength of the ISRF is calculated by Equation (\ref{eq:Qavg_num}) with $u_{\lambda}$ replaced by $u_{\lambda,\rm ISRF}$, yielding 
\bea
\bar{\lambda}&=&\frac{\sum_{i}\bar{\lambda}_{i}u_{\rad,i}}{\sum_{i}u_{\rad,i}}=\frac{0.53\cm\K\sum_{i}W_{i}T_{\star,i}^{3}}{\sum_{i}W_{i}T_{\star,i}^{4}}\nonumber\\
&=&1.28\mum,\label{eq:wmean_3star}
\ena
and, the average torque can be rewritten as
\bea
\overline{\Gamma}_{\rm RAT}=\sum_{i}\Gamma_{\rm RAT}(T_{\star,i})=\pi a^{2}
\gamma u_{\rad,\rm ISRF} \left(\frac{\bar{\lambda}}{2\pi}\right)\overline{Q}_{\Gamma},~~~
\ena
where
\bea
\overline{Q}_{\Gamma} = \frac{\sum_{i}\bar{\lambda}_{i}u_{\rad,i}\overline{Q}_{\Gamma,i}}{\bar{\lambda}u_{\rm rad,\rm ISRF}}.\label{eq:Qmean_3star}
\ena

Equations (\ref{eq:wmean_3star}) and (\ref{eq:Qmean_3star}) allow us to calculate the mean wavelength and the average RAT efficiency for an arbitrary radiation source consisting of different stars, such as star clusters. 

For the ISRF, plugging $\bar{\lambda}_{i}$, $u_{\rad,i}$, and $\overline{Q}_{\Gamma,i}$ given by Equation (\ref{eq:Qavg_star}) into the above equation, one obtains
\bea
\overline{Q}_{\Gamma}\simeq
8\left(\frac{\bar{\lambda}}{a}\right)^{-3},\label{eq:Qavg_3star}
\ena
which is valid for $a\lesssim  (Q_{\Gamma,\max}/8)^{1/3}\bar{\lambda}\approx
\bar{\lambda}/2.7$. Large grains of $a>\bar{\lambda}/2.7$ then have $\bar{Q}_{\Gamma}=Q_{\Gamma,\max}=0.4$.

\begin{figure}
\includegraphics[width=0.5\textwidth]{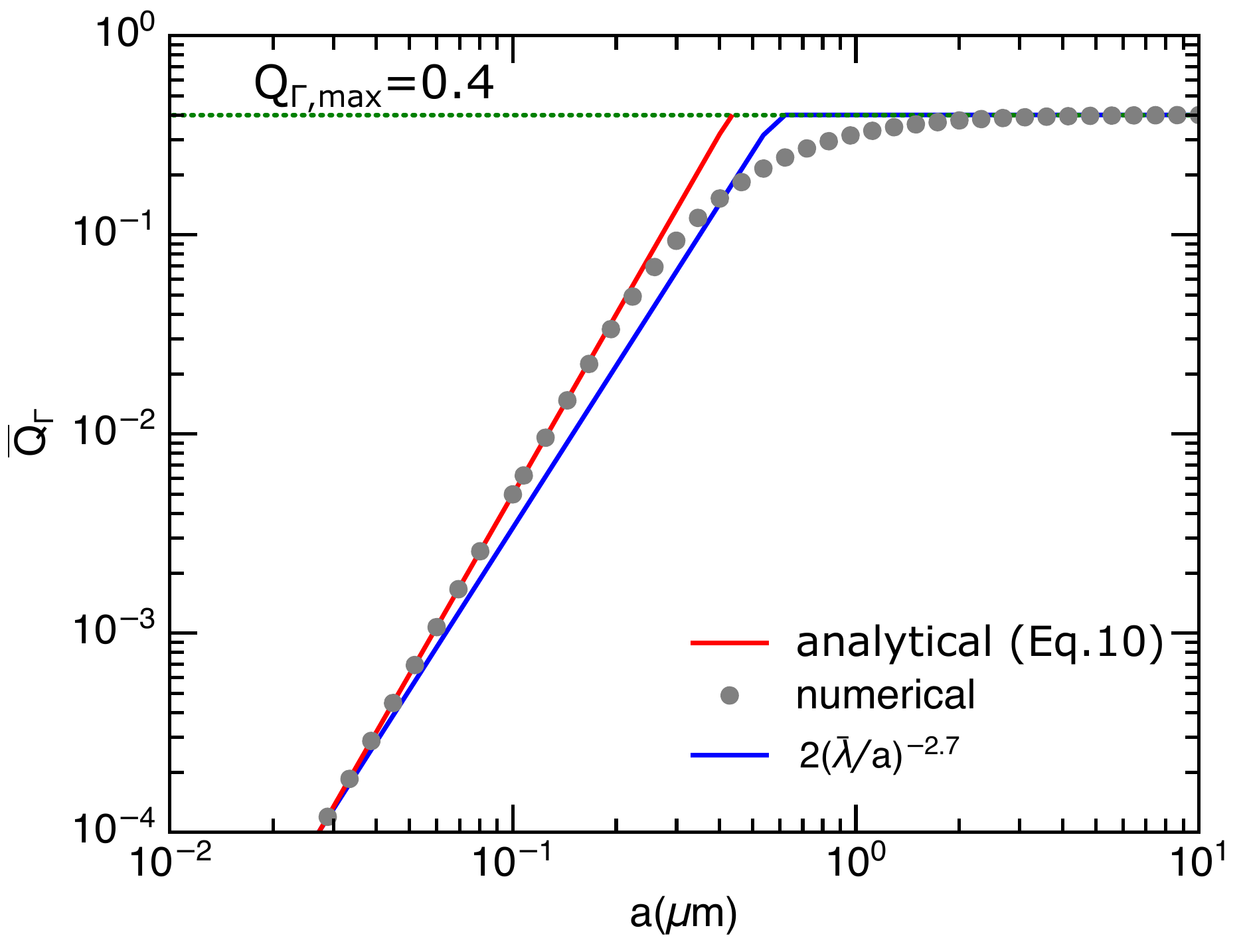}
\includegraphics[width=0.5\textwidth]{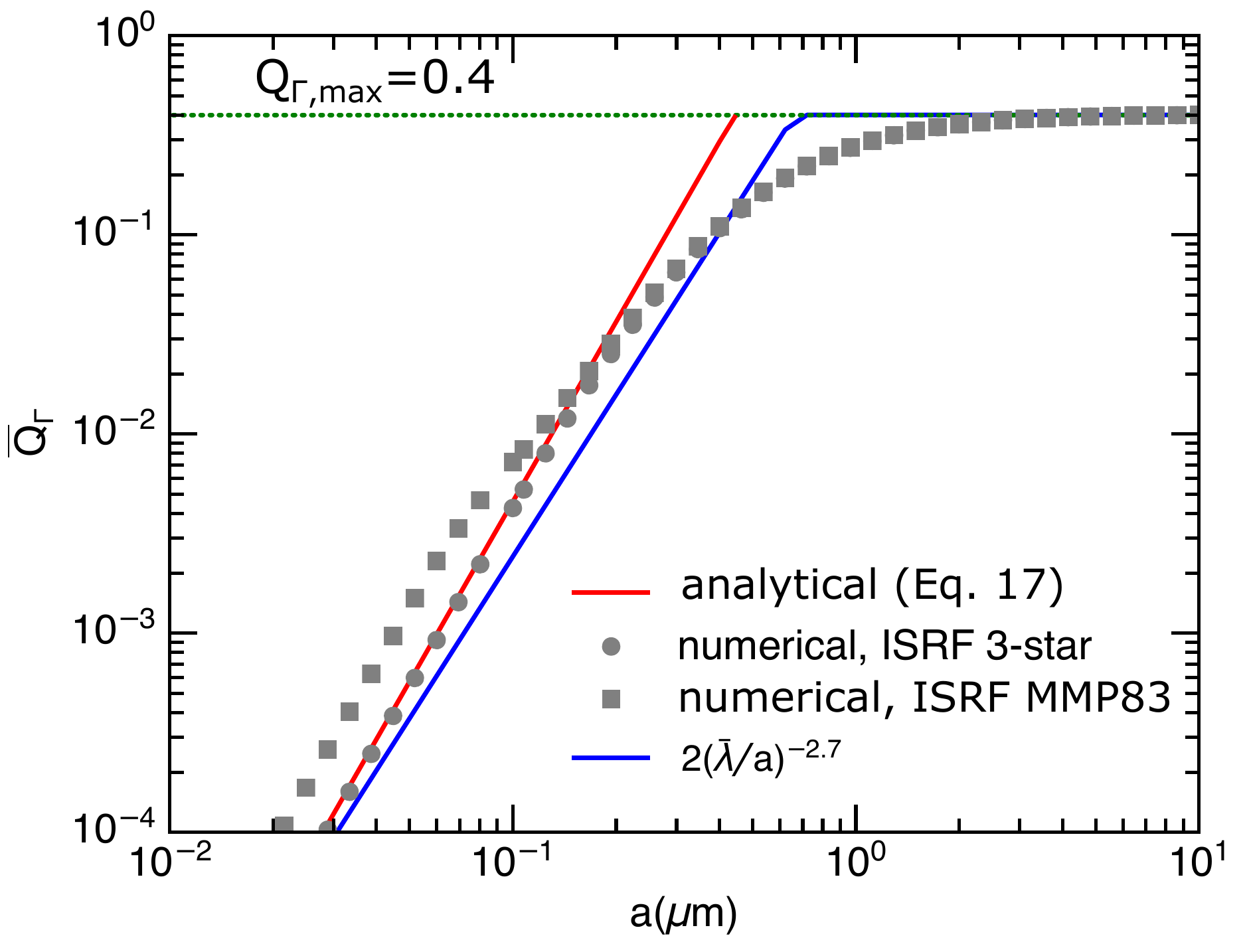}
\caption{Comparison of the average RAT efficiency obtained from analytical formulae with those from numerical calculations for a stellar radiation (upper panel) and the ISRF (lower panel). An analytical fit of $\overline{Q}_{\Gamma}=2(\bar{\lambda}/a)^{-2.7}$ from HLM14 is also shown for comparison. Analytical formulae are in excellent agreement with numerical results for the stellar spectrum and the ISRF described by three stars. For the full ISRF, the analytical result is different from numerical ones due to the contribution of the enhanced UV spectrum.}
\label{fig:QRAT}
\end{figure}

The upper panel of Figure \ref{fig:QRAT} compares the results obtained from the analytical formulae with numerical calculations using Equation (\ref{eq:Qavg_num}) for the stellar radiation spectrum of temperature $T_{\star}=5000\K$ with $\bar{\lambda}=1\mum$. For the stellar radiation, an excellent agreement is seen for $a<a_{\rm trans}=\bar{\lambda}/1.8$. For $a$ in the vicinity of $a_{\rm trans}$, the analytical result deviates significantly from the numerical result. 

The lower panel of Figure \ref{fig:QRAT} compares our analytical result with numerical results for the ISRF described by three stars (i.e., ISRF 3-star, Eq. \ref{eq:uwave_ISRF}) and the full ISRF from MMP83. For small grains of $a<a_{\rm trans}$, the analytical result agrees well with the numerical result for the ISRF 3-star but it is smaller than the numerical result for the full ISRF due to the contribution of UV radiation that is disregarded in Equation (\ref{eq:uwave_ISRF}). Note that \cite{2014ApJ...790....6H} (HLM14) adopted a power-law scaling of $\overline{Q}_{\Gamma}\simeq 2(\bar{\lambda}/a)^{-2.7}$ for $a<\bar{\lambda}/1.8$, and $\overline{Q}_{\Gamma}\sim 0.4$ for $a> \bar{\lambda}/1.8$, which satisfies the smooth transition at $a=a_{\rm trans}=\bar{\lambda}/1.8$. As shown in Figure \ref{fig:QRAT}, this scaling is in good agreement with the analytical results (within $\sim 15\%$ and $30\%$ for the stellar and ISRF radiation, respectively) agrees better with numerical results for $a\sim a_{\rm trans}$. For dense MCs, as will be shown in Section \ref{sec:GMC}, the UV radiation is rapidly attenuated due to dust absorption, therefore our analytical formulae will be in better agreement with numerical calculations.

Note that Equations (\ref{eq:Qavg_3star}) and (\ref{eq:Qavg_star}) show  that the average RAT efficiency of the ISRF is different from that of the stellar radiation by a factor of $4/3$. Therefore, in the following, we will use Equation (\ref{eq:Qavg_3star}) for deriving analytical formulae of grain alignment and disruption, unless stated otherwise.

\section{Grain Alignment and Disruption by Radiative Torques}
\label{sec:align_disrupt}
We now describe the basic theory of grain alignment and disruption, and then derive the general analytical formulae for the minimum size of aligned grains and disruption by RATs.
\subsection{Theoretical consideration}
An anisotropic radiation field can align dust grains
via the RAT mechanism (see \citealt{2007JQSRT.106..225L}; \citealt{2015ARA&A..53..501A} for reviews). For an ensemble of grains subject to only RATs, a fraction of grains is first spun-up to suprathermal rotation and then driven to be aligned with the ambient magnetic fields (so-called B-RAT) or with the anisotropic radiation direction (i.e., k-RAT) at an attractor point with high angular momentum, usually referred to as high-J attractors. The high-J attractor corresponds to the maximum angular velocity achieved by RATs. A large fraction of grains is driven to low-J attractors (\citealt{2007MNRAS.378..910L}; \citealt{2009ApJ...695.1457H}). In the presence of gas collisions, numerical simulations of grain alignment by RATs in \cite{{Hoang:2008gb},{2016ApJ...831..159H}} show that grains at the low-J attractor will be gradually transported to the high-J attractor after several gas damping times, establishing the stable efficient alignment. In dense MCs, the timescale for such a stable alignment is short compared to other dynamical timescales.
 
The existence of high-J attractors by RATs is expected for some grain shapes (\citealt{2007MNRAS.378..910L}), and it becomes universal for grains with enhanced magnetic susceptibility via iron inclusions (\citealt{2008ApJ...676L..25L}; \citealt{2016ApJ...831..159H}). This important finding is supported by simulations in \cite{2016ApJ...831..159H} for grains with various magnetic susceptibilities. Recent calculations of RATs for Gaussian random shapes by Herranen et al. (submitted) show a higher fraction of grain shapes with high-J attractors. Therefore, grain angular velocity induced by RATs is a key parameter for grain alignment (both B-RAT and k-RAT alignment) and rotational disruption.

\subsection{Maximum rotation speed by RATs}
For a radiation source with constant luminosity considered in this paper, radiative torques $\Gamma_{\rm RAT}$ is constant, and the grain angular velocity is steadily increased over time. The equilibrium rotation can be achieved at (see \citealt{2007MNRAS.378..910L}; \citealt{2009ApJ...695.1457H}; \citealt{2014MNRAS.438..680H}):
\bea
\omega_{\rm RAT}=\frac{\Gamma_{\rm RAT}\tau_{\rm damp}}{I_{a}},~~~~~\label{eq:omega_RAT0}
\ena
where $I_{a}=8\pi \rho a^{5}/15$ is the principal moment of inertia of spherical grains of radius $a$, and $\tau_{\rm damp}$ is the rotational damping time.

Total rotational damping rate by gas collisions and IR emission can be written as
\bea
\tau_{\rm damp}^{-1}=\tau_{\gas}^{-1}(1+ F_{\rm IR}),\label{eq:taudamp}
\ena
where $\tau_{\rm gas}$ is the damping time due to gas collisions followed by evaporation and given by
\bea
\tau_{\gas}&=&\frac{3}{4\sqrt{\pi}}\frac{I_{a}}{1.2n_{\rm H}m_{\rm H}
v_{\rm th}a^{4}}\nonumber\\
&\simeq& 8.3\times 10^{3}a_{-5}\hat{\rho}\left(\frac{1}{n_{3}T_{1}^{1/2}}\right)~{\rm yr},~~
\ena
where $\hat{\rho}=\rho/(3\g\cm^{-3})$ with $\rho$ being the dust mass density, $v_{\rm th}=\left(2k_{\B}T_{\rm gas}/m_{\rm H}\right)^{1/2}$ is the thermal velocity of a gas atom of mass $m_{\rm H}$ in the gas with temperature $T_{\gas}=(10\K) T_{1}$ and density $n_{\H}=(10^{3}\cm^{-3})n_{3}$, the spherical grains are assumed (\citealt{2009ApJ...695.1457H}; \citealt{1996ApJ...470..551D}).

The second term, $\tau_{\rm IR}=F_{\rm IR}\tau_{\rm gas}$ is the rotational damping due to 
IR photons emitted by the grain carry away part of the grain's angular. For strong radiation fields or not very small sizes, grains can achieve equilibrium temperature, such that the IR damping coefficient (see \citealt{1998ApJ...508..157D}) can be calculated as
\bea
F_{\rm IR}\simeq 0.038U^{2/3}a_{-5}^{-1}n_{3}T_{1}^{1/2},\label{eq:FIR}
\ena 
which is subdominant over the gas damping for starless dense clouds.

Plugging $\tau_{\rm damp}$ and $\Gamma_{\rm RAT}$ with $Q_{\rm RAT}$ from Equation (\ref{eq:Qavg_3star}) into Equation (\ref{eq:omega_RAT}), one obtains
\bea
\omega_{\rm RAT}&=& \frac{3\gamma u_{\rm rad}a\bar{\lambda}^{-2}}{1.2n_{\rm H}\sqrt{2\pi m_{\rm H}kT_{\rm gas}}}\nonumber\\
&\simeq &1.25\times 10^{6} a_{-5}\left(\frac{\bar{\lambda}}{1.2\mum}\right)^{-2}
\left(\frac{\gamma_{-1}U}{n_{3}T_{1}^{1/2}}\right)\nonumber\\
&&\times\left(\frac{1}{1+F_{\rm IR}}\right)\rad\s^{-1},\label{eq:omega_RAT}
\ena
for grains with $a\lesssim \bar{\lambda}/2.7$, and
\bea
\omega_{\rm RAT}&=&\frac{0.15\gamma u_{\rm rad}\bar{\lambda}a^{-2}}{1.2n_{\rm H}\sqrt{2\pi m_{\rm H}kT_{\rm gas}}}\nonumber\\
&\simeq& 1.08\times 10^{8}a_{-5}^{-2}\left(\frac{\bar{\lambda}}{1.2\mum}\right) \left(\frac{\gamma_{-1}U}{n_{3}T_{1}^{1/2}}\right)\nonumber\\
&&\times\left(\frac{1}{1+F_{\rm IR}}\right)\rad\s^{-1},\label{eq:omega_RAT2}
\ena
for grains with $a> \overline{\lambda}/2.7$. Here $\gamma_{-1}=\gamma/0.1$ is the anisotropy of radiation field relative to the typical anisotropy of the diffuse interstellar radiation field. 

\subsection{Grain Alignment}

Efficient alignment is achieved only when grains can rotate with angular velocity greater than its thermal value. The thermal angular velocity is
\bea
\omega_{T}&=&\left(\frac{kT_{\rm gas}}{I_{a}}\right)^{1/2}=\left(\frac{15kT_{\gas}}{8\pi\rho a^{5}}\right)^{1/2}\nonumber\\
&\simeq &5.23\times 10^{4}\hat{\rho}^{-1/2}a_{-5}^{-5/2}T_{1}^{1/2}~ \rm rad\s^{-1}.\label{eq:omega_th}
\ena

Using the suprathermal rotation threshold of $\omega_{\rm RAT}(a_{\rm align}) = 3\omega_{T}$ as in \cite{2008MNRAS.388..117H}, one obtains the minimum size of aligned grains (hereafter alignment size) as follows:
\bea
a_{\rm align}&=&\left(\frac{n_{\rm H}T_{\rm gas}}{\gamma u_{\rm rad}\bar{\lambda}^{-2}} \right)^{2/7}\left(\frac{15m_{\rm H}k^{2}}{4\rho}\right)^{1/7}\nonumber\\
&\simeq &0.055\hat{\rho}^{-1/7} \left(\frac{n_{3}T_{1}}{\gamma_{-1}U}\right)^{2/7} \nonumber\\
    &&\times \left(\frac{\bar{\lambda}}{1.2\mum}\right)^{4/7} \left(\frac{1}{1+F_{\rm IR}}\right)^{-2/7} ~\mum,\label{eq:aalign_ana}
\ena 
which implies $a_{\rm align} \sim 0.055 \mum$ for a dense cloud of $n_{\H}=10^{3}\cm^{-3}$ exposed to the local radiation field of $\gamma=0.1$, $U=1$, and $\bar{\lambda}=1.2\mum$. The alignment size increases with with increasing $n_{\rm H}$ and decreasing $U$. 

In a very dense region with low radiation strength, the alignment size becomes large and can exceed $a_{\rm trans}$. Therefore, the appropriate equation is obtained by using $\omega_{\rm RAT}$ from Equation (\ref{eq:omega_RAT2}), yielding
\bea
a_{\rm align,lg} &=& \left(\frac{n_{\rm H}T_{\rm gas}}{\gamma u_{\rm rad}\bar{\lambda}^{2}}\right)^{2}\left(\frac{90m_{\rm H}k}{\rho}\right)\nonumber\\
&\simeq&0.88\hat{\rho}^{-1} \left(\frac{n_{6}T_{1}}{\gamma_{-1} U_{-2}}\right)^{2}\left(\frac{\bar{\lambda}}{1.2\mum}\right)^{-2}\mum,\label{eq:aalign_vlg}
\ena
where $n_{6}=n_{\H}/(10^{6}\cm^{-3})$ and $U_{-2}=U/100$.

For molecular clouds, $F_{\rm IR}\ll 1$. Thus, Equation (\ref{eq:aalign_ana}) implies the alignment size when $n_{\rm H}, T_{\rm gas}$ and $T_{d}$ are known.

\subsection{Grain Rotational Disruption}

A rotating grain at angular velocity $\omega$ develops a centrifugal stress $S=\rho a^{2}\omega^{4}/4$. When the grain rotation rate is sufficiently high such as the tensile stress can exceed the tensile strength, $S_{\rm max}$, the grain is instantaneously disrupted into fragments. This mechanism is termed RAdiative Torque Disruption (RATD; \citealt{Hoang:2019da}). The critical angular velocity is obtained by setting $S=S_{\rm max}$:
\bea
\omega_{\rm disr}&=&\frac{2}{a}\left(\frac{S_{\max}}{\rho} \right)^{1/2}\nonumber\\
&\simeq& \frac{3.6\times 10^{8}}{a_{-5}}\hat{\rho}^{-1/2}S_{\max,7}^{1/2}~\rad\s^{-1},\label{eq:omega_cri}
\ena
where $S_{\max,7}=S_{\max}/(10^{7}\erg\cm^{-3})$.

The tensile strength of interstellar dust is uncertain, depending on grain structure (compact vs. composite) and composition \citep{2019ApJ...876...13H}. In dense MCs, grains are expected to be large and have composite structure as a result of coagulation process. Numerical simulations for porous grain aggregates from \cite{2019ApJ...874..159T} find that the tensile strength decreases with increasing the monomer radius and can be fitted with an analytical formula
\bea
S_{\max} &\simeq& 9.51\times 10^{4} \left(\frac{\gamma_{\rm sf}}{0.1 J m^{-2}}\right) \nonumber\\
&\times&\left(\frac{r_{0}}{0.1\mum}\right)^{-1}\left(\frac{\phi}{0.1}\right)^{1.8} \erg\cm^{-3},\label{eq:Smax}
\ena
where $\gamma_{\rm sf}$ is the surface energy per unit area of the material, $r_{0}$ is the monomer radius, and $\phi$ is the volume filling factor of monomers. For large composite grains made of monomers of radius $r_{0}=0.1\mum$ and $\phi=0.1$, Equation (\ref{eq:Smax}) implies $S_{\max}\approx 10^{5}\erg\cm^{-3}$. 

Comparing Equations (\ref{eq:omega_RAT}) and (\ref{eq:omega_cri}), one can obtain the disruption grain size:
\bea
a_{\rm disr}&=&\left(\frac{0.8n_{\rm H}\sqrt{2\pi m_{\rm H}kT_{\rm gas}}}{\gamma u_{\rm rad}\bar{\lambda}^{-2}}\right)^{1/2}\left(\frac{S_{\rm max}}{\rho}\right)^{1/4}\nonumber\\
&\simeq& 1.7 \left(\frac{n_{3}T_{1}^{1/2}}{\gamma_{-1}U}\right)^{1/2}\left(\frac{\bar{\lambda}}{1.2\mum}\right) \hat{\rho}^{-1/4}S_{\max,7}^{1/4}\nonumber\\
&&\times (1+F_{\rm IR})^{1/2}\mum,
\label{eq:adisr_ana}
\ena
which depends on the local gas properties, radiation field, and the grain tensile strength.

Due to the decrease of the rotation rate for $a>a_{\rm trans}$ (see Eq. \ref{eq:omega_RAT2}), there exist a maximum size of grains that can still be disrupted by centrifugal stress (\citealt{2020ApJ...891...38H}),
\bea
a_{\rm disr,max}&=&\frac{\gamma u_{\rm rad}\bar{\lambda}}{1.6n_{\rm H}\sqrt{2\pi m_{\rm H}kT_{\rm gas}}}\left(\frac{S_{\rm max}}{\rho}\right)^{-1/2}\nonumber\\
&\simeq& 
0.03\left(\frac{\gamma_{-1} U}{n_{3}T_{1}^{1/2}}\right)\left(\frac{\bar{\lambda}}{1.2\mum}\right)\hat{\rho}^{1/2}S_{\max,7}^{-1/2}\nonumber\\
&\times&(1+F_{\rm IR})^{-1}\mum.\label{eq:adisr_up}
\ena 

Table \ref{tab:parameters} lists the physical parameters of the local environment to calculate grain alignment and disruption by RATs, including the gas density ($n_{\H}$), radiation strength ($U$), the color of the radiation field ($\bar{\lambda}$), and the tensile strength of grain materiel ($S_{\max}$).

\begin{table}
\caption{Physical Parameters of Grain Alignment and Disruption by RATs}\label{tab:parameters}
\begin{tabular}{l l l} \hline\hline
{Parameters} & Alignment Size & Disruption Size\cr
\hline\\
Gas density& $n_{\H}$ & $n_{\H}$ \cr
Gas temperature & $T_{\gas}$ & $T_{\gas}$ \cr
Anisotropy degree & $\gamma$ & $\gamma$\cr
Radiation strength & $U$ & $U$  \cr
Mean wavelength & $\bar{\lambda}$ & $\bar{\lambda}$\cr
Tensile strength & & $S_{\max}$\cr

\hline
\hline
\end{tabular}
\end{table}

\section{Dense clouds without embedded stars}\label{sec:GMC}
We now apply the analytical formulae obtained in the previous section to study grain alignment and disruption inside a dense MC without and with an embedded source.


\subsection{Radiation Spectrum}
Here, we assume that a dense MC is only illuminated by the ISRF with $u_{\lambda,\rm ISRF}$, as described by Equation (\ref{eq:uwave_ISRF}). The radiation field inside the MC includes the attenuated ISRF (aISRF) and far-IR emission from dust grains. The aISRF dominates the radiation spectrum for $\lambda<20\mum$, whereas far-IR component dominates for $\lambda>20\mum$ \citep{1983ApJ...266..470M}.

The spectral energy density of the aISRF, at visual extinction $A_{V}$ from the surface, can be described by radiative transfer as follows:
\bea
u_{\lambda}(A_{V})=u_{\lambda}(A_{V}=0)e^{-\tau(\lambda)},\label{eq:ured}
\ena
where $\tau(\lambda)=A_{\lambda}/1.086$, and $u_{\lambda}(A_{V}=0)$ is the spectral density at the cloud surface. 

The wavelength dependence of the extinction (i.e., extinction curve), $A_{\lambda}/A_{V}$, is taken from \cite{1989ApJ...345..245C}. Here, we assume the ratio of total-to-selective extinction, $R_{V}=A_{V}/E(B-V)=4$ with the reddening $E(B-V)=A_{B}-A_{V}$, which is larger than the standard value of the diffuse ISM of $3.1$ (\citealt{2001ApJ...548..296W}) due grain growth in dense MCs.

The mean wavelength of the aISRF is computed by Equation (\ref{eq:Qavg_num}) with $u_{\lambda}(A_{V})$. The radiation density is $u_{\rm rad}=\int u_{\lambda}(A_{V})d\lambda$.

The spectral energy density of far-IR thermal dust emission, $u_{\lambda,td}$, is approximately given by (MMP83) 
\bea
\lambda u_{\lambda,td}=\frac{4\pi \lambda}{c}\sum_{i=1}^{3}W_{i}B_{\lambda}(T_{d,i}),
\ena
where $T_{d,i=1-3}=40, 24, 10\K$, and $W_{i=1-3}=2.5\times 10^{-5}, 10^{-3}, 6\times 10^{-3}$ (see Table 2, \citealt{1983A&A...128..212M}).

The attenuated ISRF component is important for alignment of dust grains of sizes $a< 10\mum$. For very large grains (e.g., $a>100\mum$), thermal emission from dust becomes important because their size is of the same order of the mean wavelength which have strong RATs.

\begin{figure}
\includegraphics[width=0.5\textwidth]{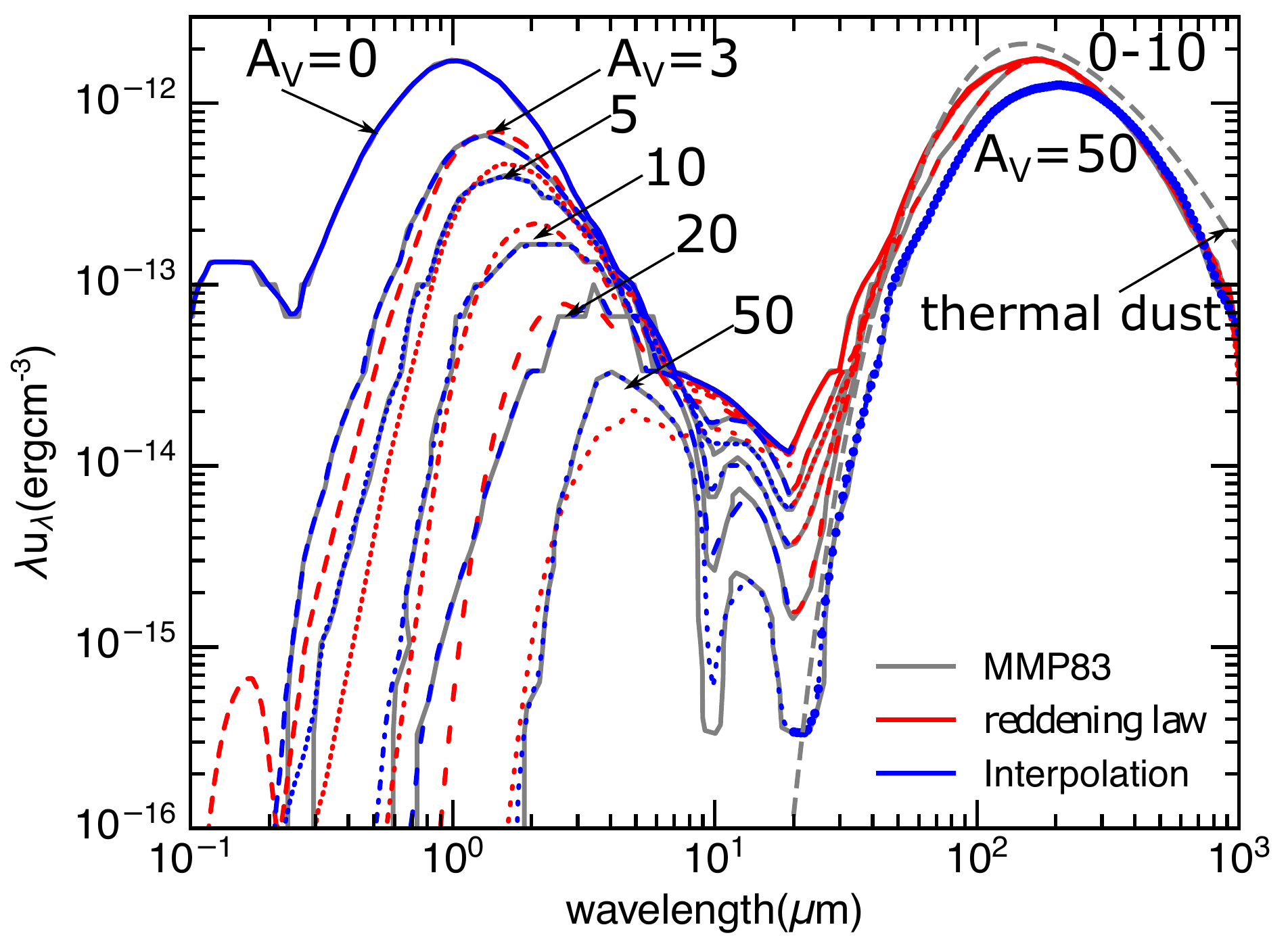}
\caption{Radiation fields at various optical depth $A_{V}$ inside a dense MC at galactocentric distance of $R_{g}=5$ kpc obtained from the reddening law with $R_{V}=4$ compared to the results from MMP83 and its interpolation. The reddening law provides good agreement with MMP83 for $\lambda<20\mum$.}
\label{fig:radfield}
\end{figure}

Figure \ref{fig:radfield} shows the spectral energy density at different visual extinction, $A_{V}$, inside a MC obtained from the reddening law (Eq. \ref{eq:ured}) with $R_{V}=4$ (red color lines), compared to the numerical results of MMP83 (filled circles) and its interpolation (blue lines). The results obtained from the reddening law are in good agreement with the numerical results, although the agreement is poorer for $A_{V}=20$ and $50$.\footnote{As already mentioned in MMP83, their radiative transfer is less accurate for $A_{V}\gtrsim 20$.} Moreover, we can see that UV-optical radiation is rapidly decreased, but NIR radiation is slowly decreased. The reason is that the optical depth at NIR is much smaller than the visual extinction. For the optical-NIR wavelengths, the extinction can be described by a power law of $A_{\lambda}/A_{V}\sim (\lambda/0.55\mum)^{-b}$ with the slope $b\sim 2$. Thus, for $A_{V}=10$, one has $A_{1\mum}=0.4A_{V}=4$ and $A_{2\mum}=0.14A_{V}=1.4$. Therefore, NIR photons are weakly absorbed and can still be sufficient to align large micron-sized grains at this large $A_{V}\sim 10$, as we show in the following section. 

When the spectral energy density is known, we calculate the radiation strength, $U$, and the mean wavelength, $\bar{\lambda}$, for different $A_{V}$. The upper and lower panel of Figure \ref{fig:U_AV} shows the decrease of $U$ and the increase of $\bar{\lambda}$ with increasing $A_{V}$, obtained using the reddened spectrum (solid line) and the spectrum from \citealt{1983A&A...128..212M} (filled circles).

To describe the decrease of $U$ with $A_{V}$ due to dust absorption, we introduce the analytical function,
\bea
U=\frac{U_{0}}{1+c_{1}A_{V}^{c_{2}}},\label{eq:U_AV}
\ena
where $U_{0}$ is the radiation strength at the cloud surface, and $c_{1},c_{2}$ are the fitting parameters. For a giant MC at D=5 kpc studied in MMP83, $U_{0}\sim 3$ (see Figure \ref{fig:U_AV}), and $U_{0}=1$ for the ISRF in the solar neighborhood. 

Assuming that the dust opacity at long wavelengths is $\kappa_{d}\propto \lambda^{-\beta}$, the equilibrium temperature of grains at $A_{V}$ is $T_{d}= T_{d,0} U^{1/(4+\beta)}$ where $T_{d,0}$ is the grain temperature at $U=1$. Through this paper, we assume $\beta=2$ for silicates and $T_{d,0}=16.4\K$ (see \citealt{2011piim.book.....D}).

Chi-square fitting of Equation (\ref{eq:U_AV}) to the numerical results obtained for the reddening law yields the best-fit parameters of $(c_{1},c_{2})=(0.42,1.22)$. As shown in the upper panel of Figure \ref{fig:U_AV}), the function provides a good fit for $A_{V}<20$, but it overestimates the numerical result by $30\%$ at $A_{V}=50$ (see dotted line vs. solid line). 

To describe the increase of the mean wavelength with $A_{V}$ due to reddening effect, we introduce the analytical function,
\bea
\bar{\lambda}=\bar{\lambda}_{0}(1+c_{3}A_{V}^{c_{4}}),\label{eq:wavemean_AV}
\ena
where $\bar{\lambda}_{0}=1.3\mum$ for the diffuse ISRF at $A_{V}=0$, $c_{3},c_{4}$ are the model parameters. Chi-square fitting of the above equation to the numerical results yield $(c_{3},c_{4})=(0.27,0.76)$. As shown in the lower panel of Figure \ref{fig:U_AV}, the analytical expression fits very well the numerical result (see dotted line vs solid line).

\begin{figure}
\includegraphics[width=0.5\textwidth]{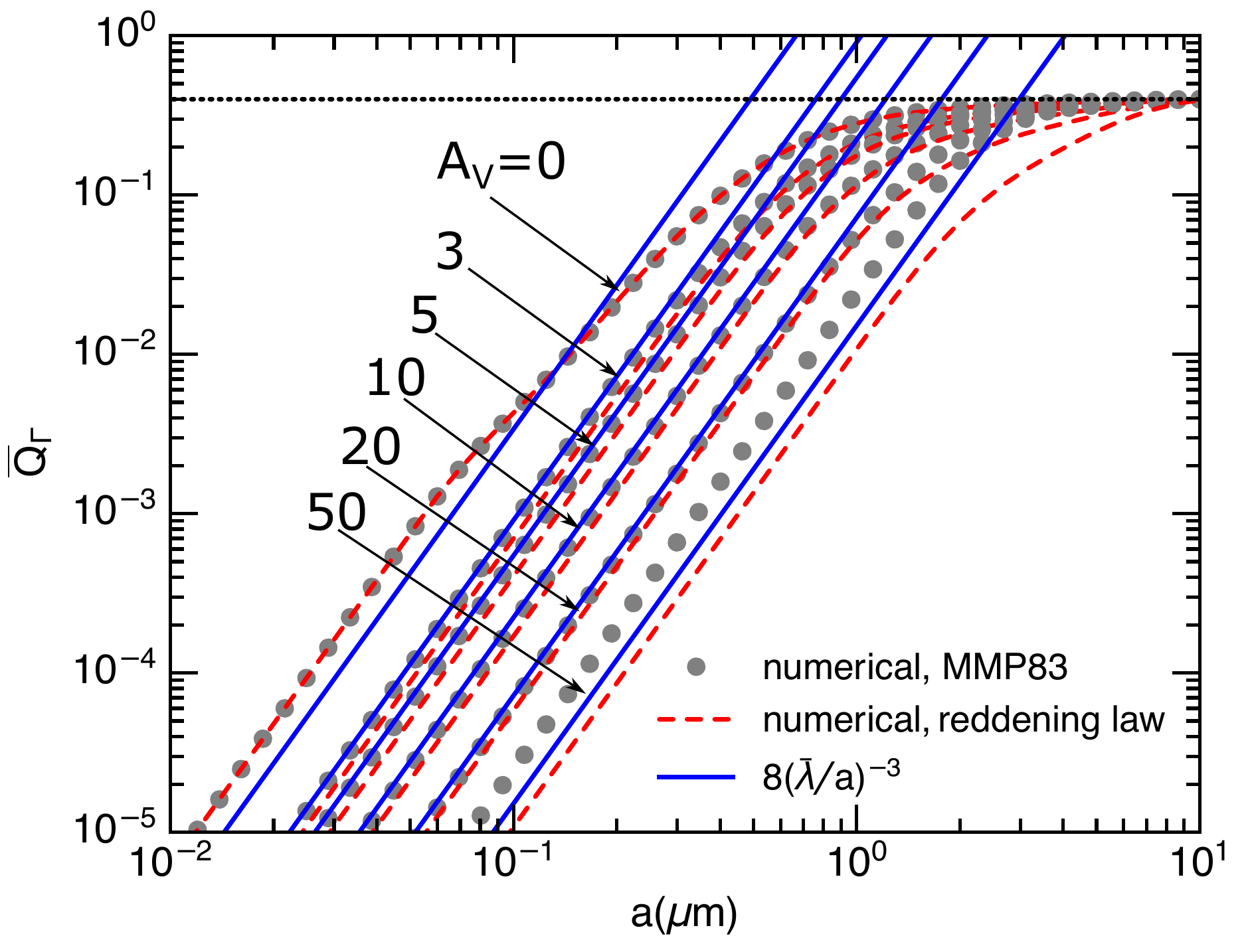}
\caption{RAT efficiency as a function of the grain size for radiation fields in a MC. The analytical approximation (Eq. \ref{eq:Qavg_3star}) fits well the numerical calculations.}
\label{fig:Qavg_GMC}
\end{figure}

Figure \ref{fig:Qavg_GMC} compares the average RAT efficiency inside a MC obtained with our analytical formula (Eq. \ref{eq:Qavg_3star}) with numerical calculations using the attenuated radiation spectrum from MMP83 and the reddening law (Eq. \ref{eq:ured}). The analytical formula which is derived for the ISRF agrees well with the numerical results for attenuated radiation fields. Therefore, one can use it for studying grain alignment by RATs in dense MCs.

\subsection{Alignment and Disruption Size}
Assuming the gas-dust thermal equilibrium, the gas temperature can also be described by a power law
\bea
T_{\gas}=\frac{T_{\rm 0}}{(1+c_{1}A_{V}^{c2})^{1/(4+\beta)}},
\ena
where $T_{0}=T_{d,0}U_{0}^{1/(4+\beta)}\K$ is the gas temperature at $A_{V}=0$. This implies a slow degree of $T_{\gas}$ with $A_{V}$ due to its dependence as a slope of $1/(4+\beta)$.

Plugging $U$ and $\bar{\lambda}$ from Equations (\ref{eq:U_AV}) and (\ref{eq:wavemean_AV}) into (\ref{eq:aalign_ana}), one obtains the analytical formula for alignment size:
\bea
    a_{\rm align} \simeq &&
       0.055\hat{\rho}^{-1/7} \left(\frac{n_{3}T_{0,1}}{\gamma_{-1} U_{0}}\right)^{2/7}\left(\frac{\bar{\lambda}_{0}}{1.2\mum}\right)^{4/7}\nonumber\\
&&\times (1+0.42A_{V}^{1.22})^{(2-2/(4+\beta))/7} (1+0.27A_{V}^{0.76})^{4/7} \mum,~~~~
\label{eq:aalign_AV_GMC}
\ena
where $T_{0,1}=T_{0}/10\K$, and the IR damping is omitted due to its subdominance. For $a>a_{\rm trans}$, only large grains can be aligned and the alignment size is given by Equation (\ref{eq:aalign_vlg}), yielding
\bea
a_{\rm align,lg}&\simeq& 0.88\hat{\rho}^{-1}\left(\frac{n_{6}T_{1}}{\gamma_{-1}U_{-2}}\right)\left(\frac{\bar{\lambda}}{1.2\mum}\right)\mum.
\ena

The grain disruption size is then given by
\bea
a_{\rm disr}&\simeq& 
2.92 \left(\frac{\bar{\lambda}_{0}}{1.2\mum}\right)(1+0.27A_{V}^{0.76})S_{\max,7}^{1/4}\nonumber\\
&&\times\left(\frac{n_{3}T_{1}^{1/2}}{\gamma_{-1}U_{0}}\right)^{1/2}(1+0.42A_{V}^{1.22})^{1/2} \mum,
\label{eq:adisr_AV_GMC}
\ena
provided that $a_{\rm disr}<\bar{\lambda}/2.7$.

Equations (\ref{eq:aalign_AV_GMC})-(\ref{eq:adisr_AV_GMC}) provide analytical results for the alignment and disruption sizes of grains at optical depth $A_{V}$ inside the starless MC, which depend on five physical parameters of the gas ($n_{\H}$) and the radiation field at the cloud surface ($\gamma, U_{0},\bar{\lambda}_{0}$).

To check the validity of these formulae, we numerically calculate the alignment size using $\omega_{\rm RAT}$ from Equation (\ref{eq:omega_RAT0}) where $\Gamma_{\rm RAT}$ is numerically computed using Equation (\ref{eq:GammaRAT_num}) for the attenuated ISRF from MMP83.

\begin{figure}
\includegraphics[width=0.5\textwidth]{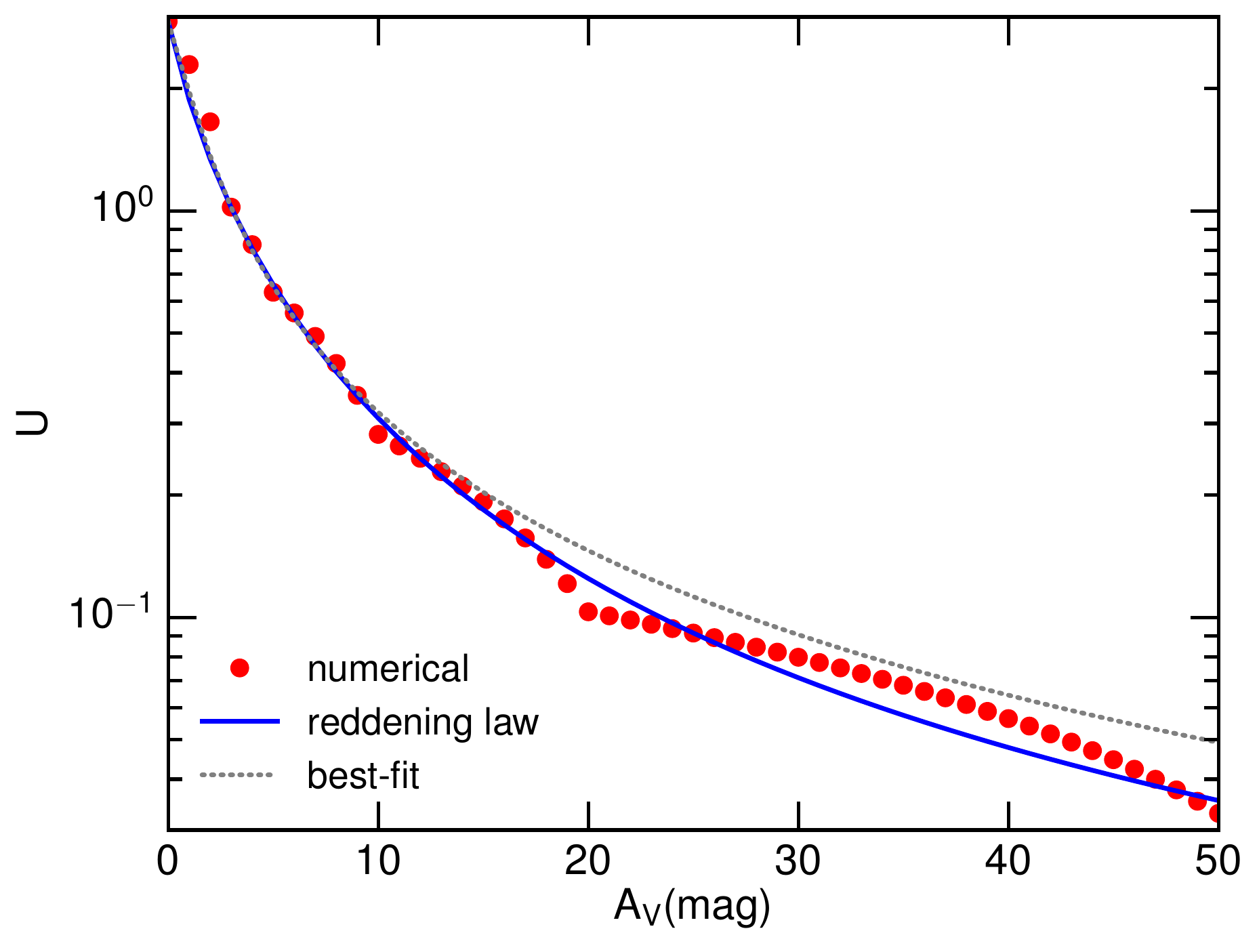}
\includegraphics[width=0.5\textwidth]{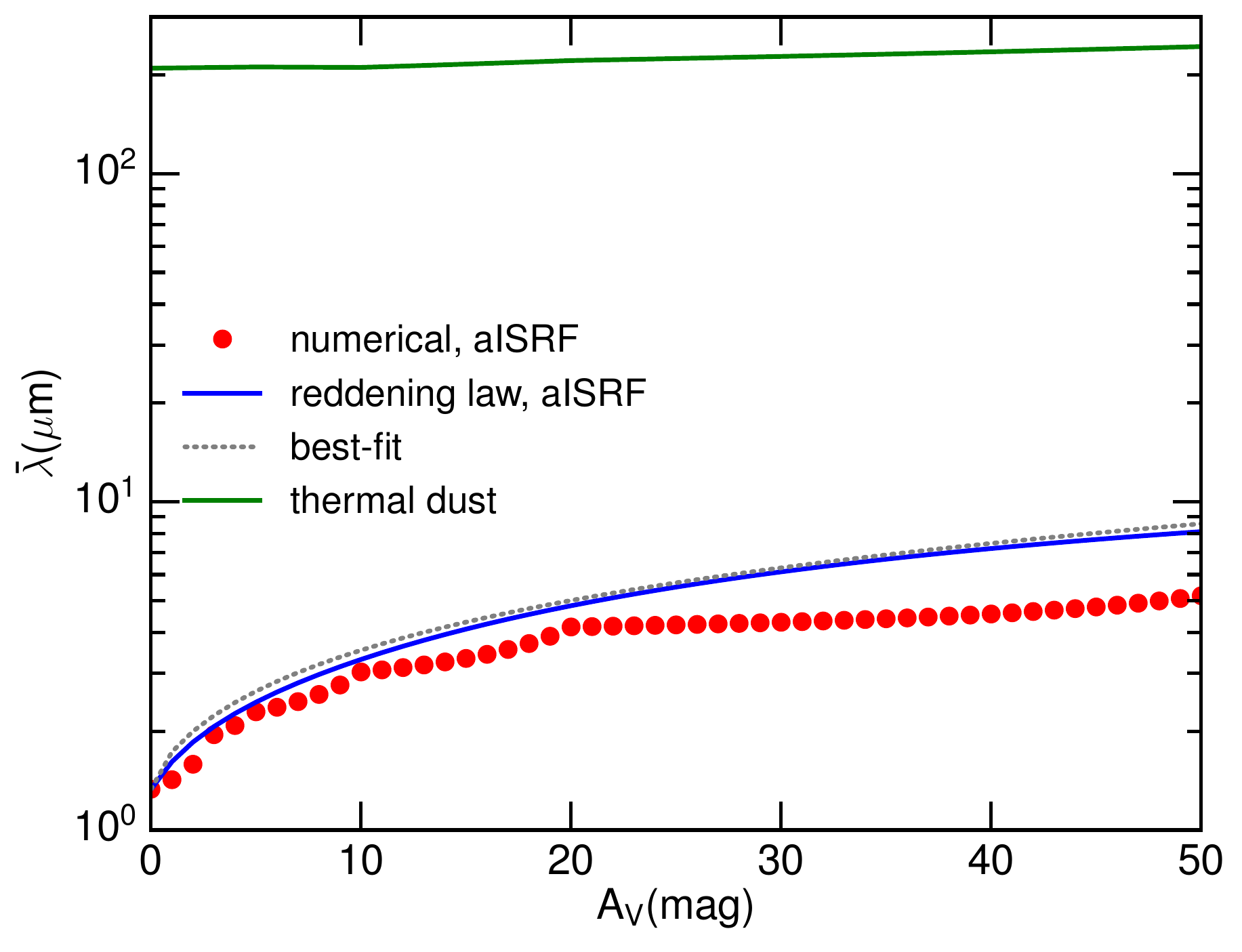}
\caption{Radiation strength (upper panel) and the mean wavelength (lower panel) as functions of the visual extinction from the surface, $A_{V}$. Symbols show numerical calculations and the solid and dotted lines show analytical results using the reddening law and best-fit to these.}
\label{fig:U_AV}
\end{figure}

\begin{figure}
\includegraphics[width=0.5\textwidth]{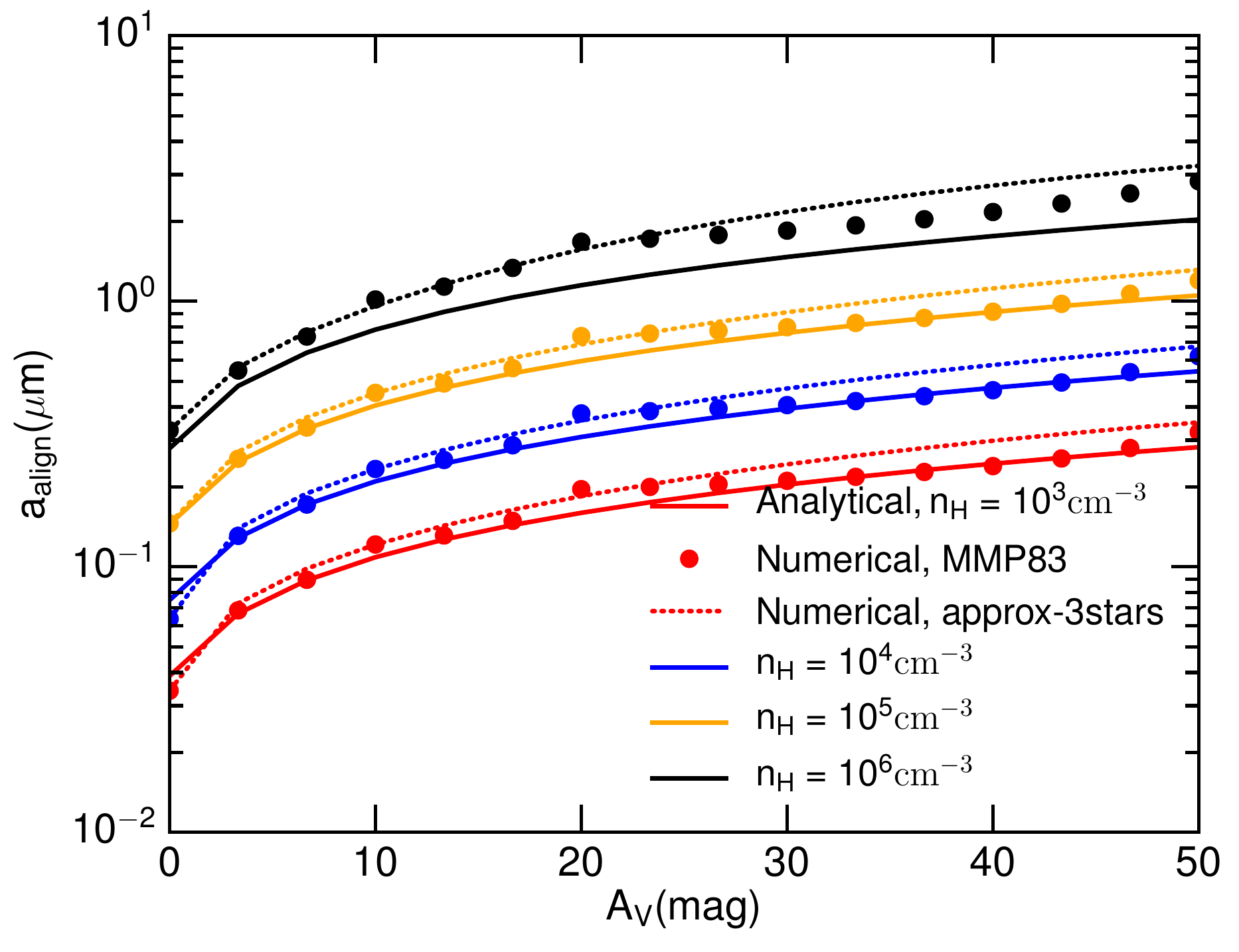}
\caption{Alignment size as a function of the surface-to-center visual extinction, $A_{V}$, obtained with our analytical result (solid lines) vs. numerical calculations (dotted and filled circle lines) for different values of gas density $n_{\rm H}$ and the anisotropy degree $\gamma=0.3$.}
\label{fig:aalign_AV}
\end{figure}

Figure \ref{fig:aalign_AV} shows $a_{\rm align}$ as a function of $A_{V}$ for analytical results (solid lines) vs. numerical results (filled circles and dotted lines), assuming the different local gas density $n_{\H}$. For numerical results, we assume the radiation field obtained from MMP83 and the reddened radiation field of the 3-star approximation, which have $U_{0}\approx 3$ and $\bar{\lambda}_{0}\approx 1.3\mum$. Our analytical results are in excellent agreement with numerical results. The alignment size increases gradually with $A_{V}$ and $n_{\H}$. For $n_{\H}=10^{4}\cm^{-3}$, standard grains (size $<0.3\mum$) can be aligned up to $A_{V}\sim 10$, and only large grains can be aligned at $A_{V}>20$ (see blue lines). For denser clouds with $n_{\H}\sim 10^{5}\cm^{-3}$, only large grains of $a>0.5\mum$ can be aligned at $A_{V}>10$, and micron-sized grains can be aligned up to $A_{V}\sim 50$ (see orange line).

\section{Dense clouds with an embedded protostar}\label{sec:protostar}
We move on to study grain alignment and disruption by internal radiation for dense clouds with an embedded source such as protostars.
\subsection{Radiation field and density profile }
Let us consider now a dense cloud with an embedded protostar of bolometric luminosity $L_{\star}$ and effective temperature $T_{\star}$. The mean wavelength of the stellar radiation spectrum, $\bar{\lambda}_{\star}$, is given by Equation (\ref{eq:wavemean_star}).

Due to the extinction by intervening dust, the radiation strength of the attenuated radiation field at radial distance $r$ from the central protostar is given by
\bea
U(r)=\frac{\int_{0}^{\infty} u_{\lambda}(T_{\star})e^{-\tau(\lambda)}d\lambda}{u_{\rm ISRF}},\label{eq:U_star_red}
\ena
where $u_{\lambda}(T_{\star})=L_{\lambda}/(4\pi r^{2}c)$ is the spectral energy density in the absence of dust extinction, $\tau_{\lambda}$ is the optical depth of intervening dust.

Following Equation (\ref{eq:wavemean_star}), the mean wavelength of the attenuated stellar spectrum is
\bea
\bar{\lambda}=\frac{\int_{0}^{\infty} \lambda u_{\lambda}(T_{\star})e^{-\tau(\lambda)}d\lambda}{\int _{0}^{\infty}u_{\lambda}(T_{\star})e^{-\tau(\lambda)}d\lambda}.\label{eq:wavemean_star_red}
\ena

The radial gas density in the envelope around the protostar can be approximately described by a power law:
\bea
n_{\H}=n_{\rm in}\left(\frac{r_{\rm in}}{r}\right)^{p}.
\ena
where $n_{\rm in}$ is the gas number density at inner radius $r_{\rm in}$. The central region of radius $r<r_{\rm in}$ is assumed to have a constant density of $n_{\rm in}$, which is implied by the fact that protostars form within a dense prestellar core. The typical slope is $p=3/2$ in the inner envelope and reaches $p=2$ in the outer envelope (\citealt{2001ApJ...547..317W}).

The central region has a column density of $N_{H,c}=n_{\rm in}r_{\rm in}$. The gas column density from the protostar at distance $r$ is given by 
\bea
N_{\rm H}(r)&=&\int_{0}^{r} n_{\rm H}(r')dr'=n_{\rm in}r_{\rm in}+\frac{n_{\rm in}r_{\rm in}}{p-1}\left[1- \left(\frac{r}{r_{\rm in}}\right)^{-p+1}\right],\nonumber\\
&\simeq &1.5\times 10^{22}n_{\rm in,8}r_{\rm in,1}\nonumber\\
&&\times\left(1+\frac{1}{p-1}\left[1-\left(\frac{r}{r_{\rm in}}\right)^{-p+1}\right]\right)\cm^{-2},
\ena
where $n_{\rm in,8}=n_{\rm in}/10^{8}\cm^{-3}$ and $r_{\rm in,1}=r_{\rm in}/10\AU$.

The wavelength-dependence dust extinction toward the central protostar is described by $A_{\lambda}=1.086\tau(\lambda)$, and the visual extinction is related to the column density as $A_{V}/N_{\rm H}=R_{V}/(5.8\times 10^{21}\cm^{-2})$ (\citealt{2011piim.book.....D}). Thus, the central region has a visual extinction of
\bea
A_{V,c}=\frac{N_{\H,c}}{5.8\times 10^{21}\cm^{-2}}R_{V}\simeq 10.3 n_{\rm in,8}r_{\rm in,1}\frac{R_{V}}{4.0}.\label{eq:AV_c} 
\ena

The visual extinction at radial distance $r$ in the envelope is then given by
\bea
A_{V,\star}(r)&=&\left(\frac{N_{\rm H}(r)}{5.8\times 10^{21}\cm^{-2}}\right)R_{V},\nonumber\\
&\simeq&A_{V,c}\left(1+\frac{1}{p-1}\left[1-\left(\frac{r}{r_{\rm in}}\right)^{-p+1}\right]\right).~~~\label{eq:AV_r}
\ena
For a steep density slope of $p=2$, the total extinction is $A_{V,\star}\approx 2A_{V,c}$. Similarly, for a shallower density profile of $p=3/2$, the total extinction is $A_{V,\star}\approx 3A_{V,c}$.

The dust extinction results in the decrease of the radiation strength and increase of the mean wavelength with visual extinction $A_{V,\star}$. Following Section \ref{sec:GMC}, the radiation strength at $A_{V,\star}$ from the source is described by
\bea
U=\frac{U_{\star}}{1+c_{1}A_{V,\star}^{c_{2}}}=\frac{U_{\rm in}}{1+c_{1}A_{V,\star}^{c_{2}}}\left(\frac{r_{\rm in}}{r}\right)^{2},\label{eq:U_AV_star}
\ena
where $U_{\star}$ is the radiation strength at radial distance $r$ in the absence of dust extinction, $U_{\rm in}=L_{\star}/(4\pi r_{\rm in}^{2}cu_{\rm ISRF})$ is the radiation strength at $r=r_{\rm in}$, and $c_{1},c_{2}$ are the fitting parameters. The mean wavelength of the attenuated stellar spectrum is
\bea
\bar{\lambda}=\bar{\lambda}_{\star}(1+c_{3}A_{V,\star}^{c_{4}}),\label{eq:wavemean_AV_star}
\ena
where $c_{3}$ and $c_{4}$ are the fitting parameters.

Assuming the gas-dust thermal equilibrium and using Equation (\ref{eq:U_AV}), one obtains the gas temperature as a power-law
\bea
T_{\gas}=T_{\rm in}\left(\frac{r_{\rm in}}{r}\right)^{q}(1+c_{1}A_{V,\star}^{c2})^{-q/2},
\ena
where $T_{\rm in}=T_{d,0}U_{\rm in}^{1/(1+\beta)}\K$ is the grain temperature at $r_{\rm in}$ and $q=2/(1+\beta)$. 

\begin{figure}
\includegraphics[width=0.5\textwidth]{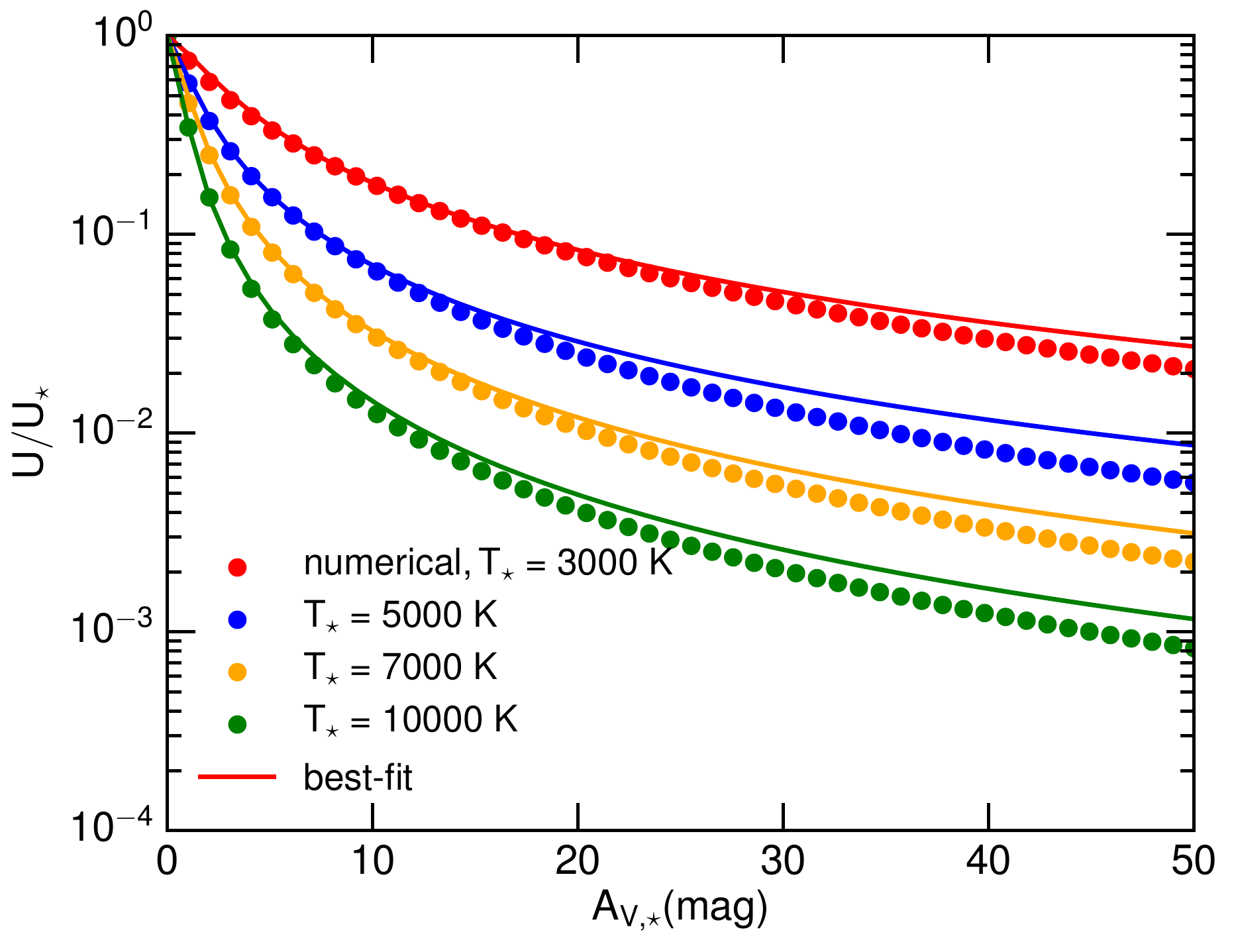}
\includegraphics[width=0.5\textwidth]{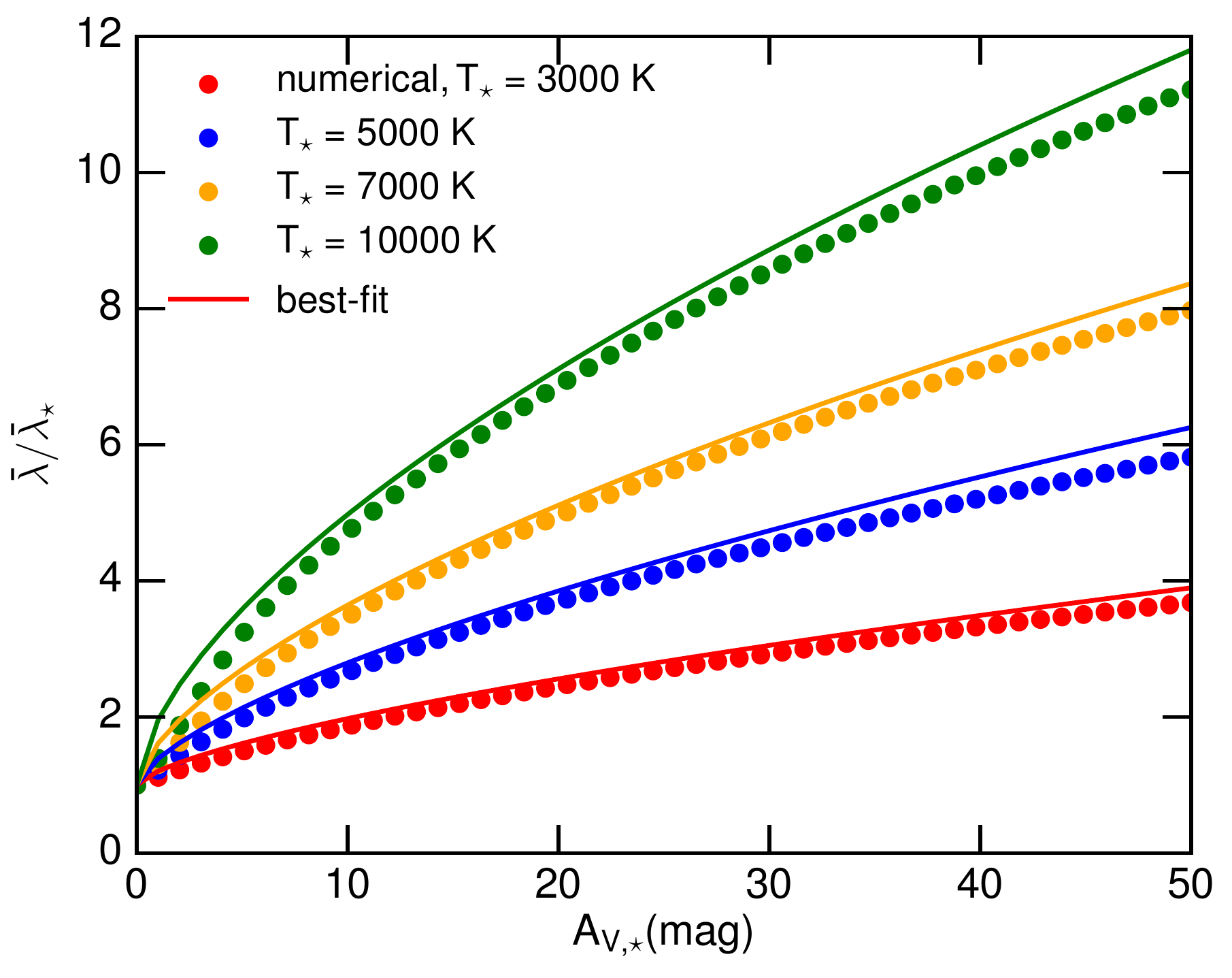}
\caption{Variation of the normalized radiation strength 
($U/U_{\star}$) and normalized mean wavelength ($\bar{\lambda}/\bar{\lambda}_{\star}$) as functions of $A_{V,\star}$ from the central protostar where $U_{\star}$ and $\bar{\lambda}_{\star}$ are the strength and mean wavelength of the stellar radiation spectrum in the absence of dust extinction. Various temperatures of protostars are considered. Best-fit to the numerical results are shown in solid lines.}
\label{fig:U_AV_star}
\end{figure}

Figure \ref{fig:U_AV_star} shows the variation of $U$ and $\bar{\lambda}$ with $A_{V,\star}$ obtained from numerical calculations using Equations (\ref{eq:U_star_red}) and (\ref{eq:wavemean_star_red}) for the different stellar temperatures. The best-fit models obtained from Equations (\ref{eq:U_AV_star}) and (\ref{eq:wavemean_AV_star}) with the best-fit parameters listed in Table \ref{tab:coefficient} are also shown in the solid lines for comparison. The fit is better for lower $T_{\star}$. The reason is that higher $T_{\star}$ experiences faster attenuation due to larger UV energy that have larger extinction, resulting in a steeper variation of $U$ and $\bar{\lambda}$ on $A_{V}$. Nevertheless, the difference is less than $\sim 50\%$ only for $A_{V}<50$.

\begin{table}
\caption{Best-fit parameters of the radiation strength and mean wavelength to numerical calculations for different stellar temperatures.}\label{tab:coefficient}
\begin{tabular}{l l l l l} \hline\hline
{$T_{\star}(\K)$} & $c_{1}$ & $c_{2}$  & $c_{3}$  & $c_{4}$ \cr
\hline\\
3000 & 0.25& 1.26& 0.21&  0.67 \cr
5000 & 0.61& 1.33& 0.39& 0.66 \cr
7000 & 1.07& 1.40& 0.60& 0.64 \cr
10000 & 1.83 & 1.51& 0.90& 0.63\cr
\cr
\hline
\hline
\end{tabular}
\end{table}

\subsection{Alignment and disruption}
Plugging $n_{\H}, T_{\gas}$, $U$, and $\bar{\lambda}$ into Equation (\ref{eq:aalign_ana}) one obtains the alignment size at optical depth $A_{V,\star}$ from the protostar,
\bea
    a_{\rm align}\simeq 
    &&0.028\hat{\rho}^{-1/7} \left(\frac{U_{\rm in,6}}{n_{\rm in,8}T_{\rm in,2}}\right)^{-2/7}\left(\frac{\bar{\lambda}_{\star}}{1.2\mum}\right)^{4/7} \nonumber\\
    &&\times (1+c_{1}A_{V,\star}^{c_{2}})^{(2-q)/7}  (1+c_{3}A_{V,\star}^{c_{4}})^{4/7}\nonumber\\
&&\times \left(\frac{r}{r_{\rm in}}\right)^{2(2-p-q)/7} \mum,
\label{eq:aalign_AV_star}
\ena
where $\gamma=1$ is adopted for the unidirectional field from the protostar, and the IR damping is omitted because $F_{\rm IR}<1$. 

Equation (\ref{eq:adisr_ana}) implies the disruption size at optical depth $A_{V}$:
\bea
a_{\rm disr}&\simeq& 
0.3 S_{\max,7}^{1/4}\left(\frac{U_{\rm in,6}}{n_{\rm in,8}T_{\rm in,2}^{1/2}}\right)^{-1/2}\left(\frac{\bar{\lambda}_{\star}}{1.2\mum}\right)\nonumber\\
&&\times(1+c_{1}A_{V,\star}^{c_{2}})^{(1-q/4)/2}(1+c_{3}A_{V,\star}^{c_{4}})\nonumber\\
&&\times\left(\frac{r}{r_{\rm in}}\right)^{(2-p-q/2)/2}\mum,
\label{eq:adisr_AV_star}
\ena
for $a_{\rm disr}<\bar{\lambda}/2.5$.

The maximum size of grain disruption is given by Equation (\ref{eq:adisr_up}),
\bea
a_{\rm disr,max}&\simeq& 0.9\hat{\rho}^{1/2}S_{\max,7}^{-1/2}\left(\frac{U_{\rm in,6}}{n_{\rm in,8}T_{\rm in,2}^{1/2}}\right)\left(\frac{\bar{\lambda}_{\star}}{1.2\mum}\right)\nonumber\\
&&\times(1+c_{1}A_{V,\star}^{c_{2}})^{(1-q/4)}(1+c_{3}A_{V,\star}^{c_{4}})\nonumber\\
&&\times\left(\frac{r}{r_{\rm in}}\right)^{(2-p-q/2)}~\mum.\label{eq:adisr_up_star}
\ena 

For the central region of $r< r_{\rm in}$, the gas density is $n_{\H}=n_{\rm in}$. Thus, the alignment and disruption size are described by Equations (\ref{eq:aalign_AV_star}) and (\ref{eq:adisr_AV_star}) by plugging $p=0$.

\begin{figure}
\includegraphics[width=0.5\textwidth]{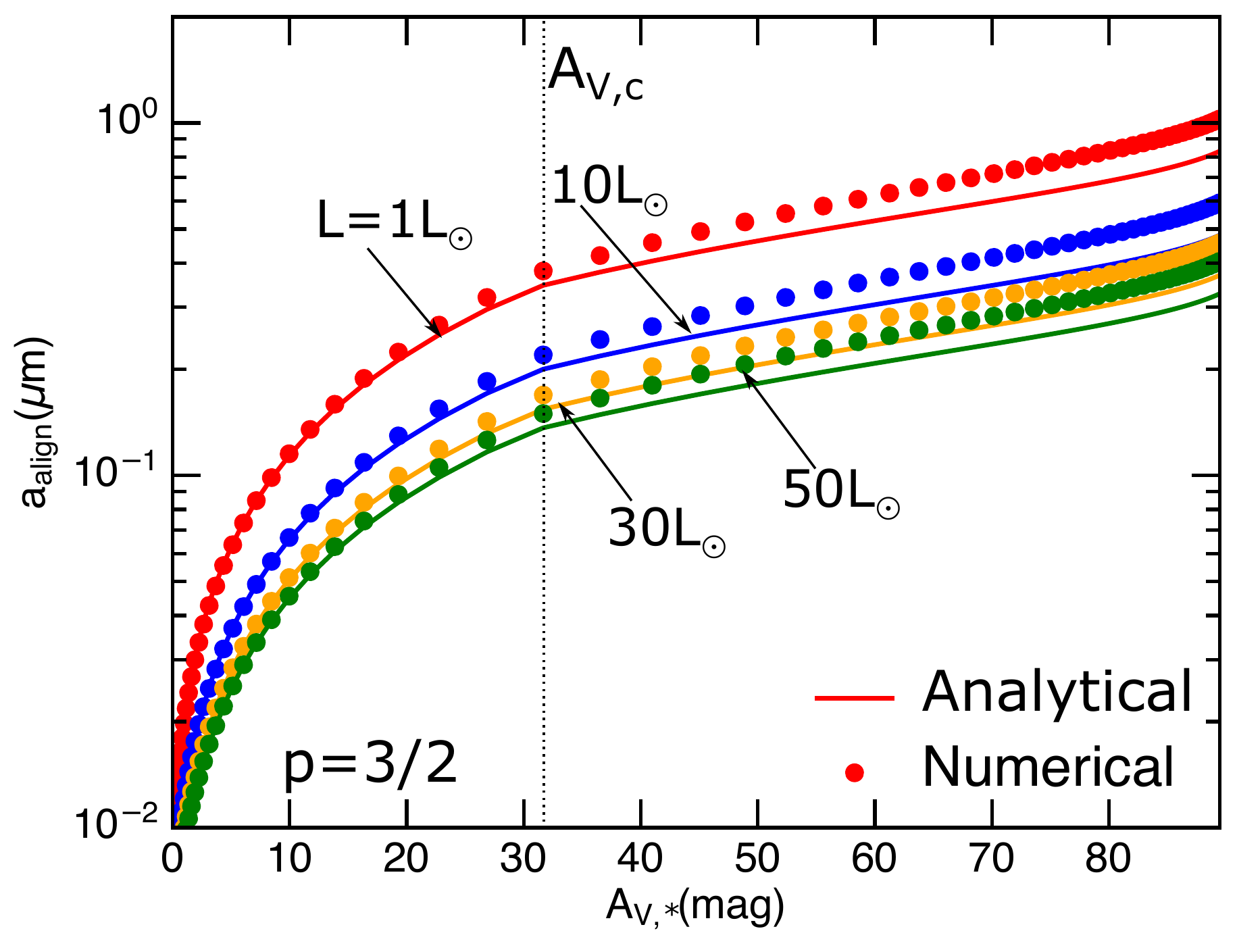}
\includegraphics[width=0.5\textwidth]{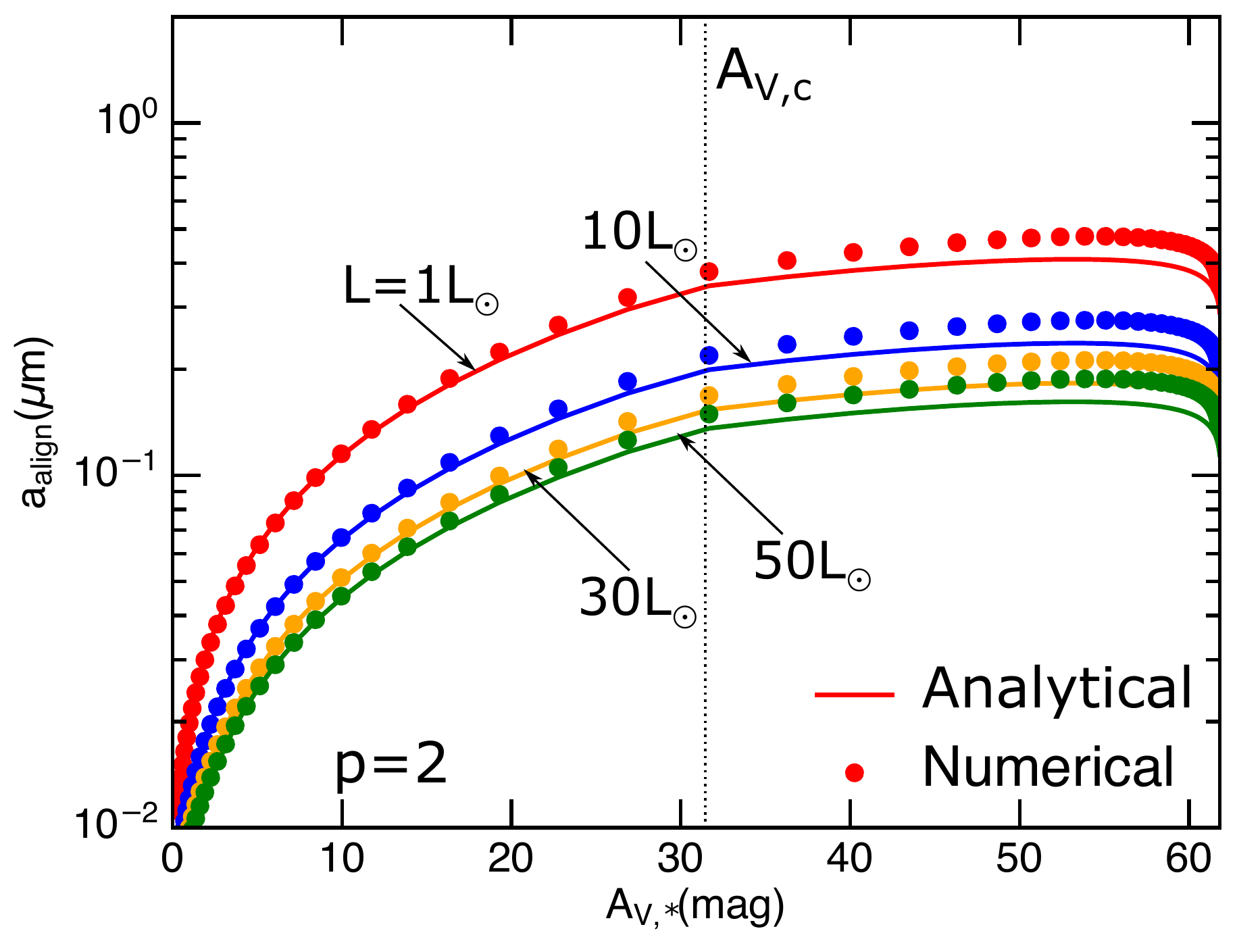}
\caption{Variation of the alignment size as a function of the visual extinction to a low-mass protostar, $A_{V,\star}$, for different luminosity $L$ using analytical formulae, assuming the radial density profile of $p=1.5$ (upper panel) and $p=2$ (lower panel). Vertical dotted line at $A_{V,c}$ marks the transition from the central core with constant density to the envelope when the density drops rapidly. Numerical results are shown by filled circles for comparison. The protostar luminosity of $L_{\star}=1-50L_{\odot}$ and temperature of $T_{\star}=3000\K$ are assumed. Good agreement between analytical and numerical results are observed.}
\label{fig:aali_AV_star_low}
\end{figure}

\begin{figure}
\includegraphics[width=0.5\textwidth]{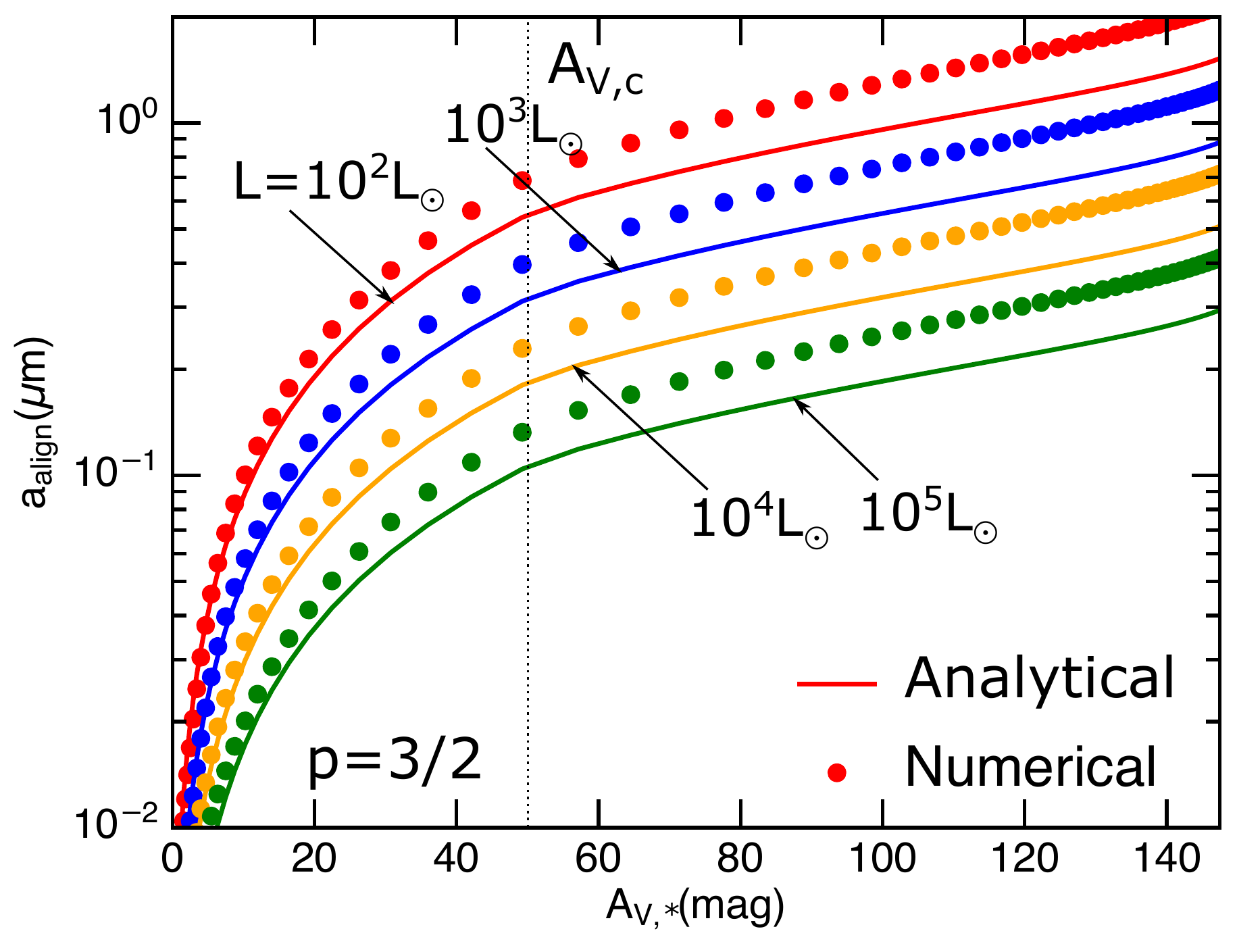}
\includegraphics[width=0.5\textwidth]{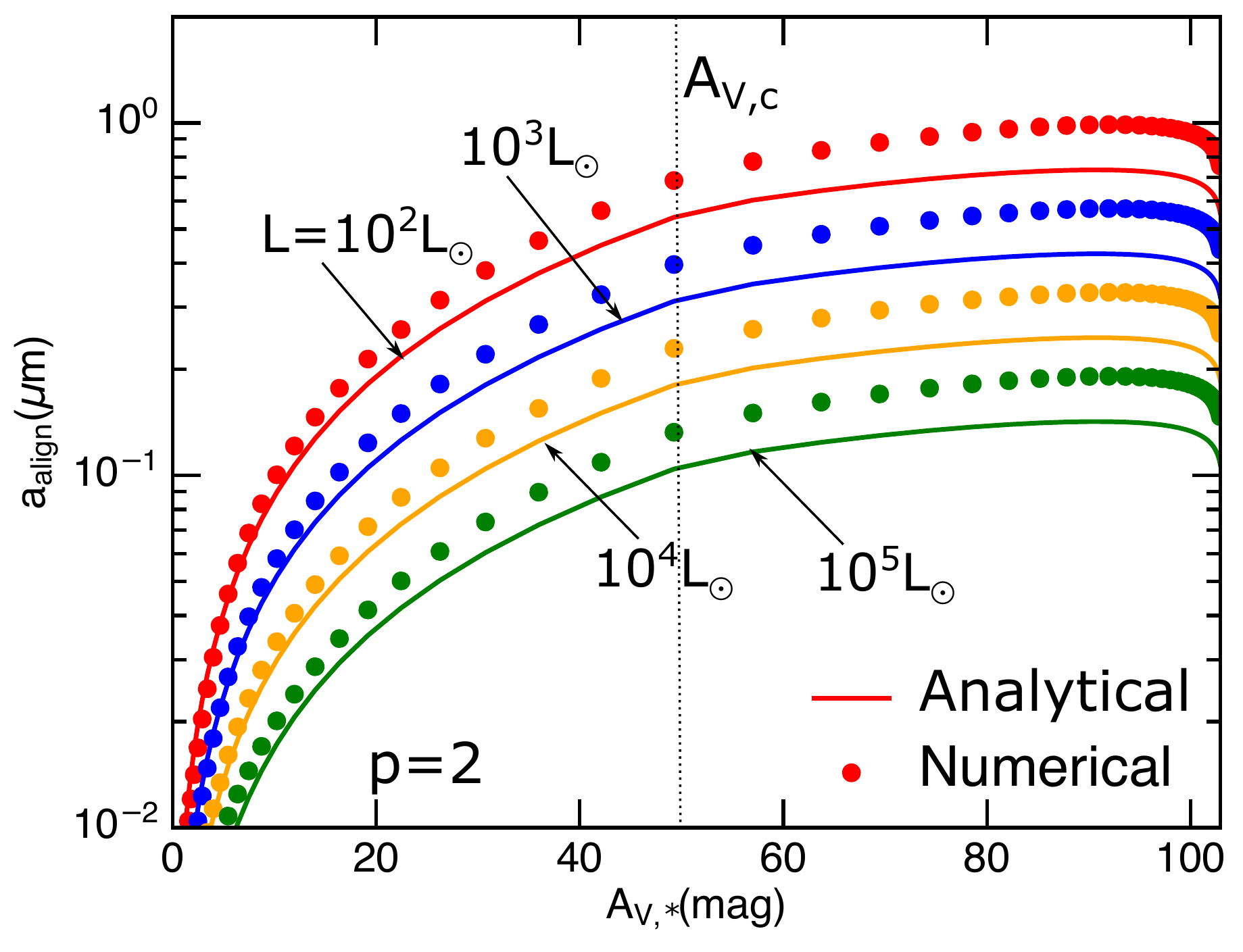}
\caption{Same as Figure \ref{fig:aali_AV_star_low} but for high-mass protostars of luminosity $L_{\star}=10^{2}-10^{5}L_{\odot}$ and $T_{\star}=10^{4}\K$.}
\label{fig:aali_AV_star_high}
\end{figure}

\begin{figure}
\includegraphics[width=0.5\textwidth]{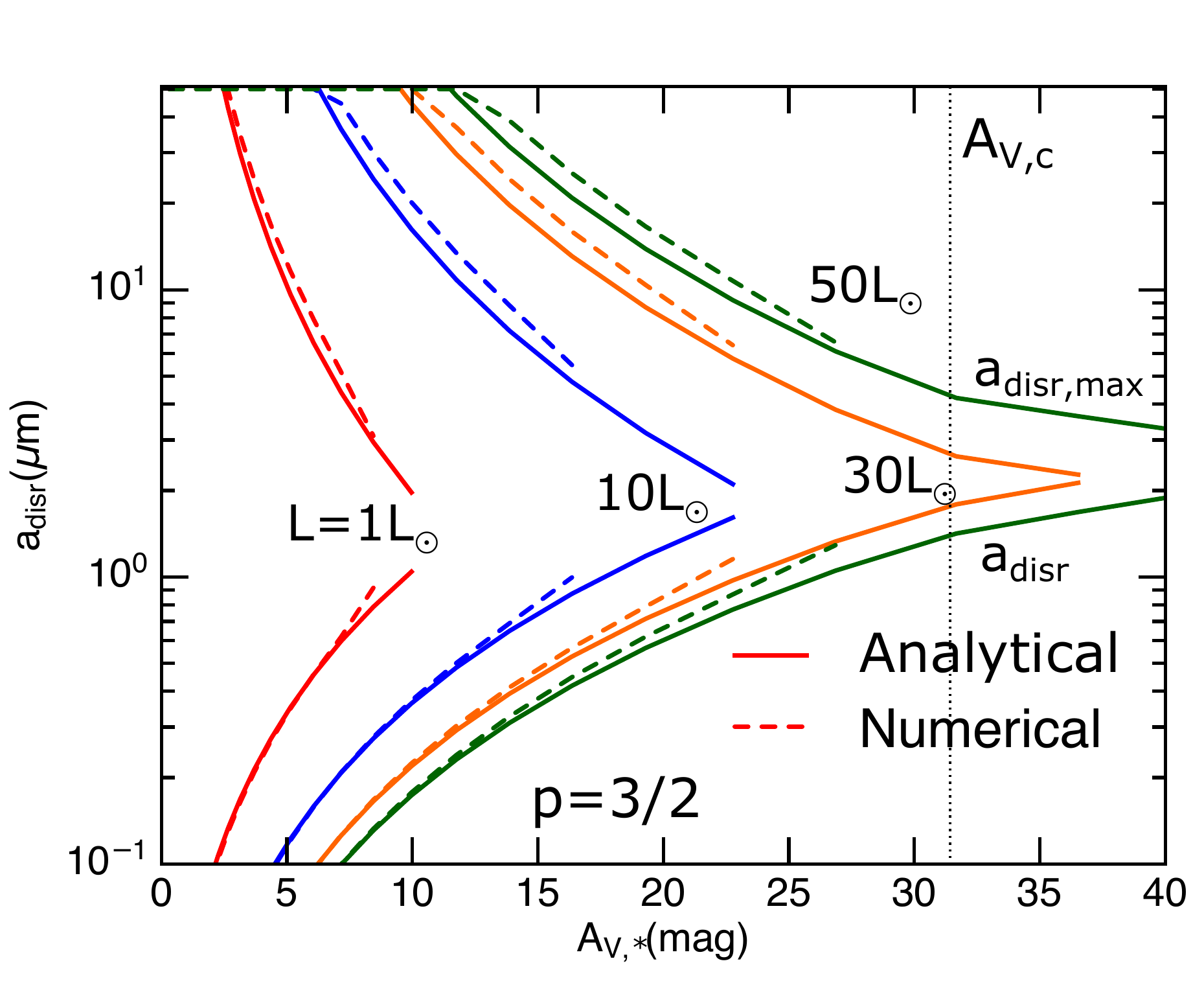}
\includegraphics[width=0.5\textwidth]{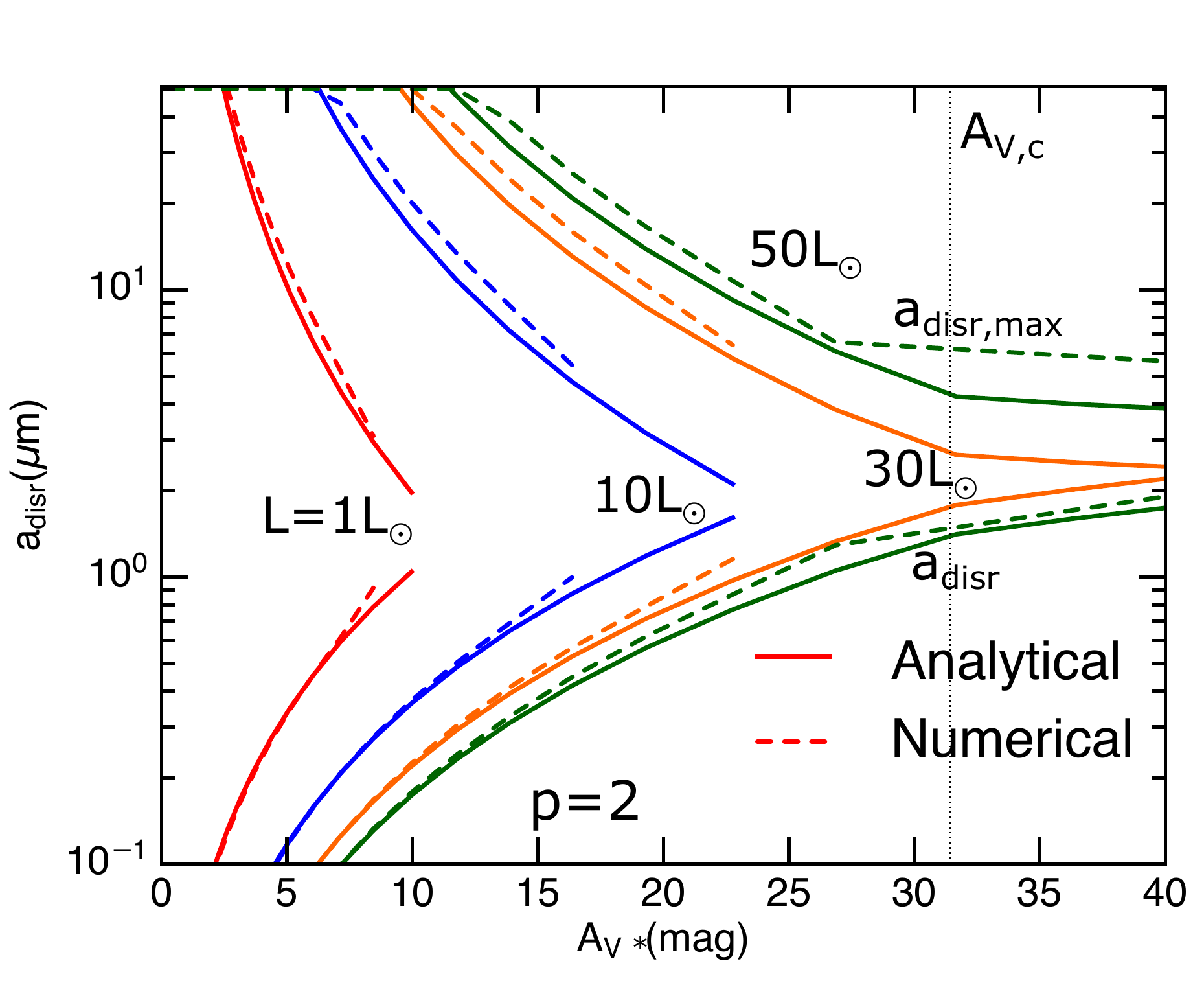}
\caption{Variation of the disruption size as a function of visual extinction for the density profile $p=3/2$ (upper) and $p=2$ (lower), for low-mass protostars with temperature of $T_{\star}=3000\K$. Disruption occurs within the central region of $A_{V}<A_{V,c}$.}
\label{fig:adisr_AV_star_low}
\end{figure}

\begin{figure}
\includegraphics[width=0.5\textwidth]{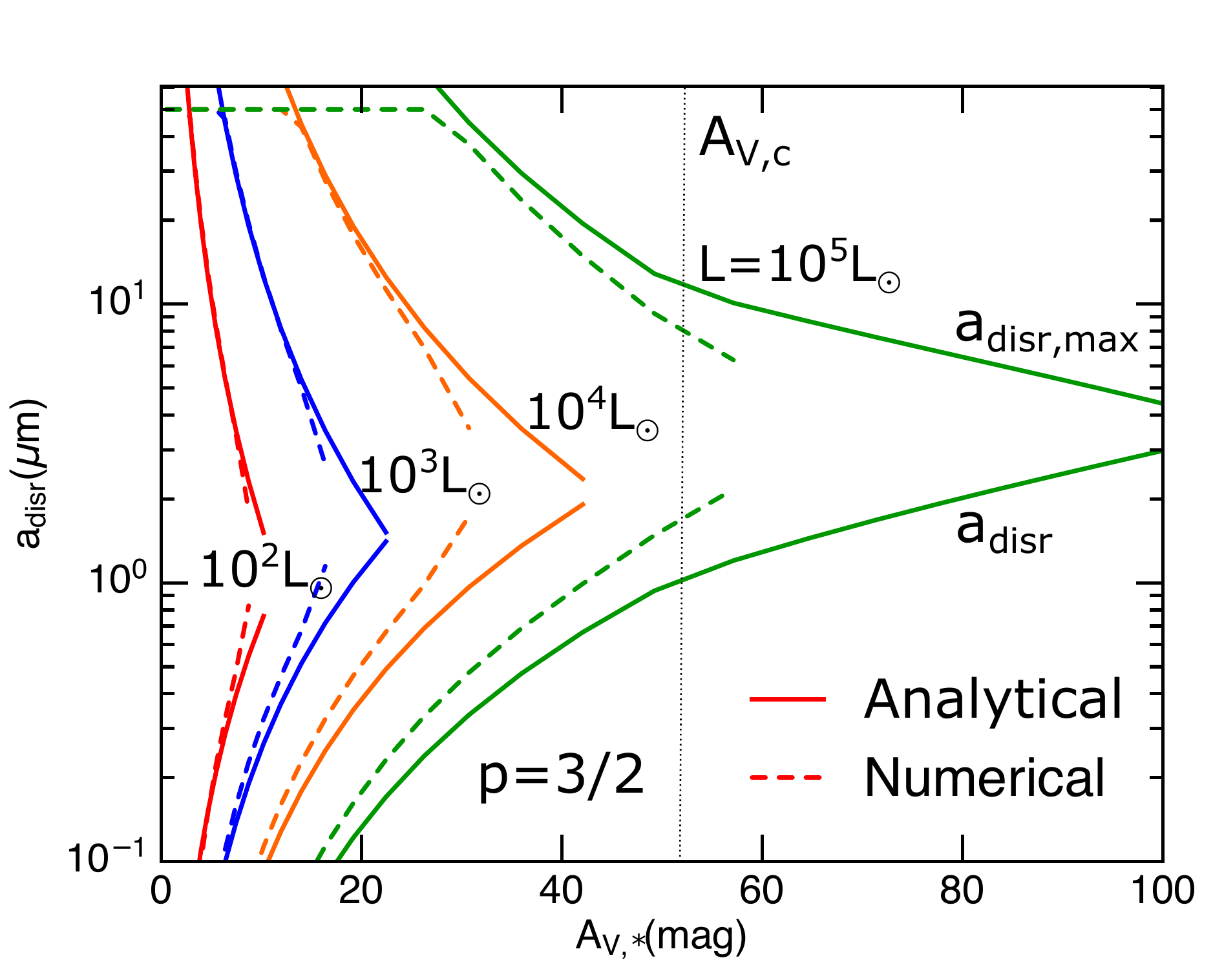}
\includegraphics[width=0.5\textwidth]{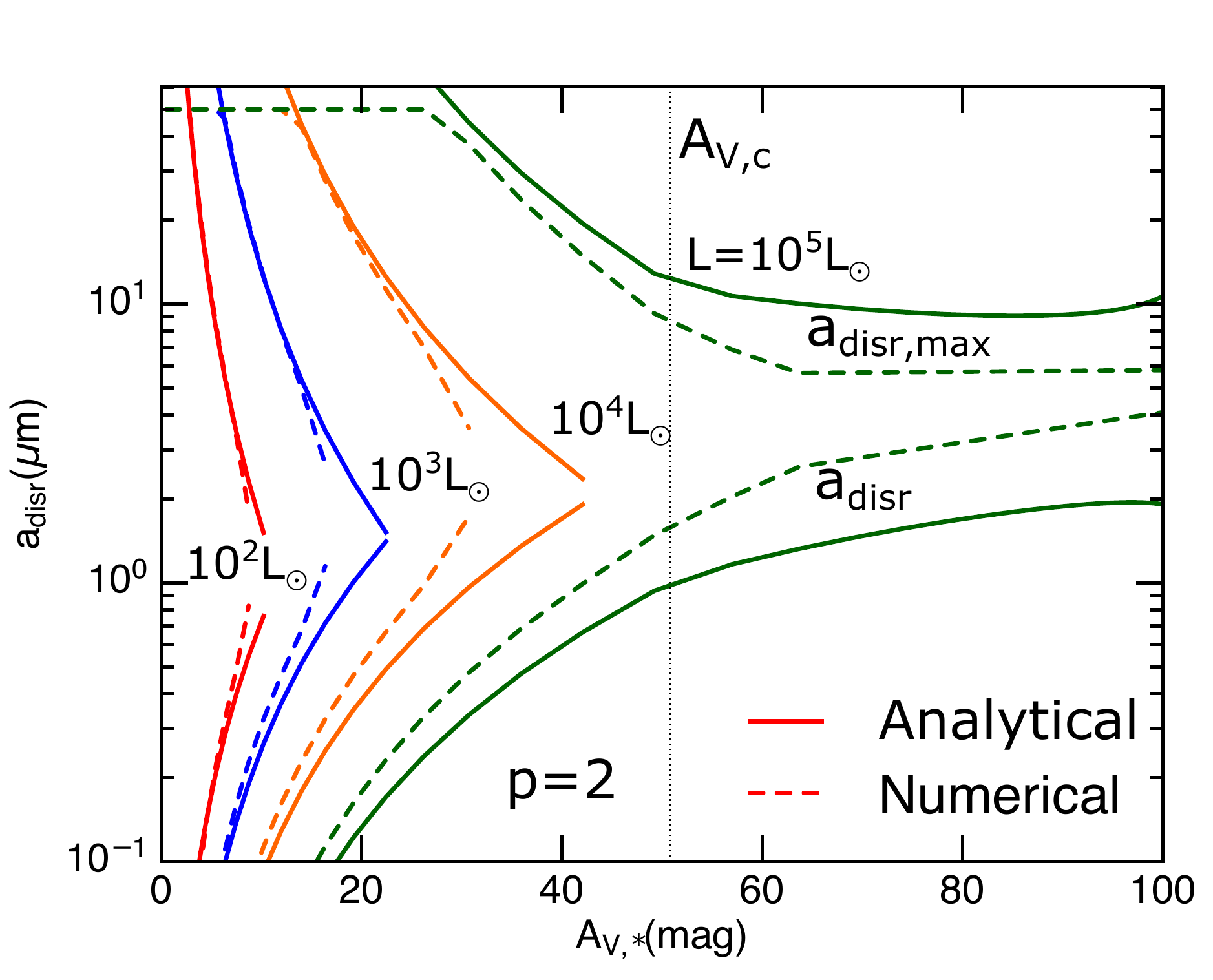}
\caption{Same as Figure \ref{fig:adisr_AV_star_low} but for high-mass protostars of temperature $T_{\star}=10^{4}\K$. Different values of the luminosity are considered. RATD mainly occurs within the central core, but it can occur in the envelope for the case of $L=10^{5}L_{\odot}$.}
\label{fig:adisr_AV_star_high}
\end{figure}

For our numerical results, we assume $n_{\rm in}=10^{8}\cm^{-3},R_{\rm in}=30$ AU for low-mass protostars and $n_{\rm in}=10^{7}\cm^{-3},R_{\rm in}=500$ AU for high-mass protostars. The luminosity of protostars is varied. The central core then has visual extinction of $A_{V,c}\approx31$ and $A_{V,c}\approx 52$ for the low-mass and high-mass protostars, respectively, assuming $R_{V}=4$.

Figure \ref{fig:aali_AV_star_low} shows the variation of $a_{\rm align}$ obtained from our analytical formula with numerical results as a function of $A_{V,\star}$ for low-mass protostars of different luminosity, assuming the typical density slope of $p=3/2$ (upper panel) and $p=2$ (lower panel). First, one can see that analytical results (solid lines) are in good agreement with numerical results (filled circles). Second, the alignment size is lower for more luminous sources. For the slope $p=1/2$, at large $A_{V,\star}>A_{V,c}$ (i.e., in the protostellar envelope), the alignment size decreases gradually with decreasing $A_{V}$, and only large grains of $a>a_{\rm align}\sim 0.15\mum$ can be aligned because of the increase of the gas density. When entering the central core of $A_{V,\star}<A_{V,c}$, the alignment size decreases rapidly with $A_{V,\star}$. For the slope $p=2$, the alignment size changes slightly in the envelope of $A_{V,\star}>A_{V,c}$ because the ratio $U/n_{\H}$ is constant as a result of both the gas density and radiation strength vary as $1/r^{2}$. When entering the central core of constant density, the alignment size decreases rapidly due to the increase of $U/n_{\H}\propto U$, similar to the slope of $p=3/2$.

Figure \ref{fig:aali_AV_star_high} shows the similar results as Figure \ref{fig:aali_AV_star_low} but for high-mass protostars. Same as low-mass protostars, only large grains of $a>0.1\mum$ can be aligned in the envelope, but small grains could be aligned in the central core due to stronger radiation. The good agreement between analytical and numerical results is seen for $A_{V,\star}<40$, and the difference is within $20\%$ for $A_{V,\star}>40$. The reason is that the analytical fit of the RAT efficiency and the mean wavelength is not good for the high stellar temperature at very large $A_{V,\star}$ (see Figure \ref{fig:Qavg_GMC}). 

Figure \ref{fig:adisr_AV_star_low} shows the variation of $a_{\rm disr}$ and $a_{\rm disr,max}$ obtained from our analytical formulae (solid lines) with numerical results (dashed lines) as function of $A_{V,\star}$ for low-mass protostars with $p=3/2$ (upper panel) and $p=2$ (lower panel), assuming the typical tense strength of $S_{\max}=10^{5}\erg\cm^{-3}$ for large composite grains of $a>0.1\mum$ (see Eq. \ref{eq:Smax}). Grain disruption mostly occurs within the central core where the radiation field is strong enough to spin grains to the critical rotation rate. Analytical and numerical results are in good agreement. The disruption zone, constrained by $a_{\rm disr}-a_{\rm disr,max}$, is more extended for more luminous sources. Therefore, grains of $a>0.1\mum$ are depleted in the central core due to the RATD effect. 

Figure \ref{fig:adisr_AV_star_high} shows the similar results as Figure \ref{fig:adisr_AV_star_low} but for high-mass protostars. For the case of $L=10^{5}L_{\odot}$, the disruption zone can extend beyond the central region into the envelope.


\section{Discussion}\label{sec:discuss}
\subsection{Comparison to previous studies on grain alignment in dense clouds}
Modeling of grain alignment by RATs for starless cores were previously presented in \cite{2005ApJ...631..361C} and \cite{2007ApJ...663.1055B}. Both studies only consider the alignment of grains by attenuated ISRF and consider the surface-to-center optical depth up to $A_{V}=10$ only. 

In this paper, we study grain alignment in dense clouds that can have large $A_{V}$. By describing the incident radiation field with two local parameters, radiation strength ($U$) and mean wavelength ($\bar{\lambda}$), we obtain the empirical relationships between $U, \bar{\lambda}$ and the visual extinction $A_{V}$. Using the RAT alignment theory, we derive a general formula for the minimum size of aligned grains in a starless MC that depends on five physical parameters, including two local parameters $(n_{\rm H},A_{V})$ and two radiation field parameters at the cloud surface ($U_{0},\bar{\lambda}_{0}$) (see Eq. \ref{eq:aalign_AV_GMC}). 

We also derive an analytical formula for the alignment size of dust grains for MCs with a central protostar. For the first time, we study the disruption of grains in dense MCs using the newly discovered RATD effect. We find that, toward the protostar, both the minimum size of grain alignment and disruption decreases, which corresponds to the increase in efficiency of grain alignment and disruption. We test our analytical formulae with numerical calculations and obtain good agreement for dense MCs. Therefore, the obtained formulae can be used to predict grain alignment and disruption, and interpret polarimetric observations from star-forming regions. 

\subsection{Polarization holes in starless dense cores}
Polarization hole, i.e., the decrease of the polarization fraction with visual extinction ($A_{V}$) or column density ($N_{H}$) was frequently observed toward starless dense cores (e.g., \citealt{2014A&A...569L...1A}). Such a decrease is described by a power law of $P_{\rm abs}/A_{V}\propto A_{V}^{-\zeta}$ for polarization of background starlight and $P_{\rm em}\propto I_{\rm em}^{-\zeta}$ for the polarization of thermal dust emission where the power slope $\zeta\sim 0-1$. 

The minimum size of aligned grains $a_{\rm align}$ in starless clouds increases with increasing $A_{V}$, as given by Equation (\ref{eq:aalign_AV_GMC}) and shown in Figure \ref{fig:aalign_AV}). Therefore, if the maximum grain size distribution is constant within the cloud, the degree of dust polarization by dichroic extinction is expected to decrease with increasing $A_{V}$ with the slope of $\zeta<1$. This successfully reproduces the popular polarization hole in starless cores. In particular, when the minimum size of aligned grains is larger than the maximum size of grains, the slope is steep with $\zeta=1$, as predicted in \cite{2005ApJ...631..361C} and \cite{2008ApJ...674..304W}. In this case, it is considered as "ideal" polarization hole, which is previously observed in starless cores (e.g., \citealt{2014A&A...569L...1A}; \citealt{2015AJ....149...31J}; \citealt{Liu:2019jf}).

We now estimate the maximum optical depth that grain alignment still exits in a starless core for a given maximum size of grains, $a_{\rm max}$. For $A_{V}\gg 1$, Equation (\ref{eq:aalign_AV_GMC}) yields
\bea
a_{\rm align} \simeq &&0.02\hat{\rho}^{-1/7} A_{V}^{5.48/7} \left(\frac{n_{3}T_{1}}{\gamma_{-1}U_{0}}\right)^{2/7} \nonumber\\ 
&&\times \left(\frac{\bar{\lambda}_{0}}{1.2\mum}\right)^{4/7} \mum.    \label{eq:aalign_loss}
\ena

Therefore, the ideal polarization hole is produced if $a_{\rm align}>a_{\rm max}$, which implies
\bea
A_{V,\rm max} &\simeq& 24.7\hat{\rho}^{1/5.48}\left(\frac{a_{\rm max}}{0.25\mum}\right)^{7/5.48} \left(\frac{\gamma_{-1}U_{0}}{n_{3}T_{1}}\right)^{1/2.74}\nonumber\\
&&\times\left(\frac{\bar{\lambda}_{0}}{1.2\mum}\right)^{-1/1.37}.\label{eq:AVmax}
\ena

For dense cores with $n_{\rm H}=10^{4},10^{5},10^{6}\cm^{-3}$, the above equation implies $A_{V,\rm max}\approx 15.9,6.9,3.0$ for the standard maximum size of $a_{\rm max}=0.25\mum$. For a larger value of $a_{\rm max}=0.5\mum$, one has $A_{V,\rm max}\approx 38.6,16.7, 7.1$ respectively, assuming the same $\gamma=0.3$. 

When the grain temperature can be inferred from the spectral energy density (SED) observations, one can also check whether large grains can be aligned in the region. Using Equation (\ref{eq:aalign_ana}), one obtains the minimum strength of the local radiation field required for alignment of largest grains by setting $a_{\rm align}=a_{\rm max}$, yielding
	\bea
	U_{\rm min}=3.8\times 10^{-2}\gamma_{-1}n_{6}T_{1}\left(\frac{a_{\rm max}}{1\mum}\right)^{-7/2}\left(\frac{\bar{\lambda}_{0}}{1.2\mum}\right)^{2},~~~~\label{eq:Umin}
	\ena    
converting to the minimum grain temperature,
\bea
T_{d,\rm min}&=&T_{d,0}U_{\rm min}^{1/6}\simeq 9.5\gamma_{-1}^{1/6}(n_{6}T_{1})^{1/6}\nonumber\\
&\times&\left(\frac{a_{\rm max}}{1\mum}\right)^{-7/12}\left(\frac{\bar{\lambda}_{0}}{1.2\mum}\right)^{1/3} \K.\label{eq:Tdmin}
\ena 

The minimum radiation strength (temperature) decreases with increasing the maximum grain size and decreasing the gas density. In a very dense region of $n_{H}\gtrsim 10^{6}\cm^{-3}$, the radiation field must be sufficiently strong so that grains can be heated to $T_{d}\gtrsim 9.5\K$ to align micron-sized grains. 


\subsection{Polarization holes toward dense clouds with an embedded source}
Polarization of dust emission in far-IR/submm (\citealt{Hildebrand:2000p6056}) is unique to probe grain alignment in the densest part of a MC thanks to the powerful polarimetric capability of the JCMT/POL-2, SOFIA/HAWC+, and ALMA. Numerous observations of dust polarization toward protostars show the decrease of the polarization fraction with the intensity as $P_{\rm em}\propto I_{em}^{-\zeta}$ with different slopes $\zeta$ for the different clouds. For instance, the value of $\zeta\sim 0.5-0.8$ is reported to low-mass (\citealt{Soam:2019fv}; \citealt{Coude:2019kq}; \citealt{2019ApJ...879...25K}) and high-mass protostars (\citealt{Liu:2020wh}). \cite{2019ApJ...880...27P} obtained the slope of $\zeta\approx 0.34$ for the Ophiuchus A cloud with an embedded source. 

Our results obtained for a cloud with an embedded source show that the alignment size $a_{\rm align}$ first changes slowly in the envelope and then decreases rapidly when entering the central region around the protostar due to the increase of the radiation flux and constant gas density (see Figures \ref{fig:aali_AV_star_low} and \ref{fig:aali_AV_star_high}). Therefore, if the grain size distribution is constant, the polarization of thermal dust emission would increase toward the central protostar, producing a slope of $\zeta=0$. However, we found that grains are rotationally disrupted by RATs in the central region (see Figure \ref{fig:adisr_AV_star_low} and \ref{fig:adisr_AV_star_high}). The removal of the largest grains by RATD is predicted to reduce the polarization at long wavelength (\citealt{2020ApJ...896...44L}), which should decrease toward the protostar, and one expects $\zeta\sim 0-1$.

It is worth to mention that \cite{Pillai:2020df} report a slope of $\zeta=0.55$ for the region with embedded stars. Their detailed modeling of polarized emission using the RAT theory could reproduce approximately the observed slope for the lower range of $I_{\rm em}$. However, their predicted polarization degree appears to increase with intensity for ${\rm log}_{10}I_{\rm em}>0.5 ~\rm Jy/pixel$, inconsistent with the observed trend (see their Figure 4b). We expect that such a discrepancy would be resolved if the RATD effect is taken into account. 

Our above discussion disregarded the effect of magnetic field fluctuations and turbulent structures, which are usually referred to explain the polarization hole (see e.g., \citealt{2014ApJS..213...13H}; \citealt{2019FrASS...6...15P}). To identify the primary process responsible for the polarization hole, it requires detailed numerical modeling of dust polarization with grain alignment and disruption for realistic three-dimensional magnetic fields combined with observational data.

\subsection{Constraining Grain Sizes in Dense Clouds}
Grains in dense MCs are expected to be larger than interstellar grains with the upper cutoff of $a_{\max}\sim 0.25-0.3\mum$ due to
grain growth as a result of accretion of gas species on to the grain surface and grain-grain collisions. Theoretical calculations predict grain growth to micron size in dense clouds (\citealt{2013MNRAS.434L..70H}). Numerous observations for dense prestellar cores show evidence of micron-sized grains through coreshine, i.e., a core become visible through scattered light in mid-infrared (e.g., \citealt{2010Sci...329.1622P}; \citealt{Lefevre:2020fw}).

Based on our analysis of grain alignment for starless dense cores, if dust polarization is still detected at large $A_{V}$, then grain growth must occur so that $a>a_{\rm align}$.\footnote{Indeed, simulations of polarized emission by dust grains aligned by RATs in Bok globules in \cite{Brauer:2016gh} confirm that the grain growth is needed to explain the observed polarization.} Various observations of dust polarization reveal existence of grain alignment in starless cores at larger extinction of $A_{V}\sim 20$ (\citealt{2017ApJ...849..157W}; \citealt{2018ApJ...868...94K}; \citealt{2019ApJ...880...27P}), which is evident of grain growth as implied from Equation (\ref{eq:AVmax}).

Equation (\ref{eq:AVmax}) shows that the maximum extinction of grain alignment increases with $a_{\max}$. Therefore, if grain growth occurs in dense MCs, one expects the alignment even at large $A_{V}$. The existence of aligned large grains in dense MCs produces the polarization at large $A_{V}$, changing the slope of the polarization fraction at larger $A_{V,\max}$. 

\begin{figure}
\includegraphics[width=0.5\textwidth]{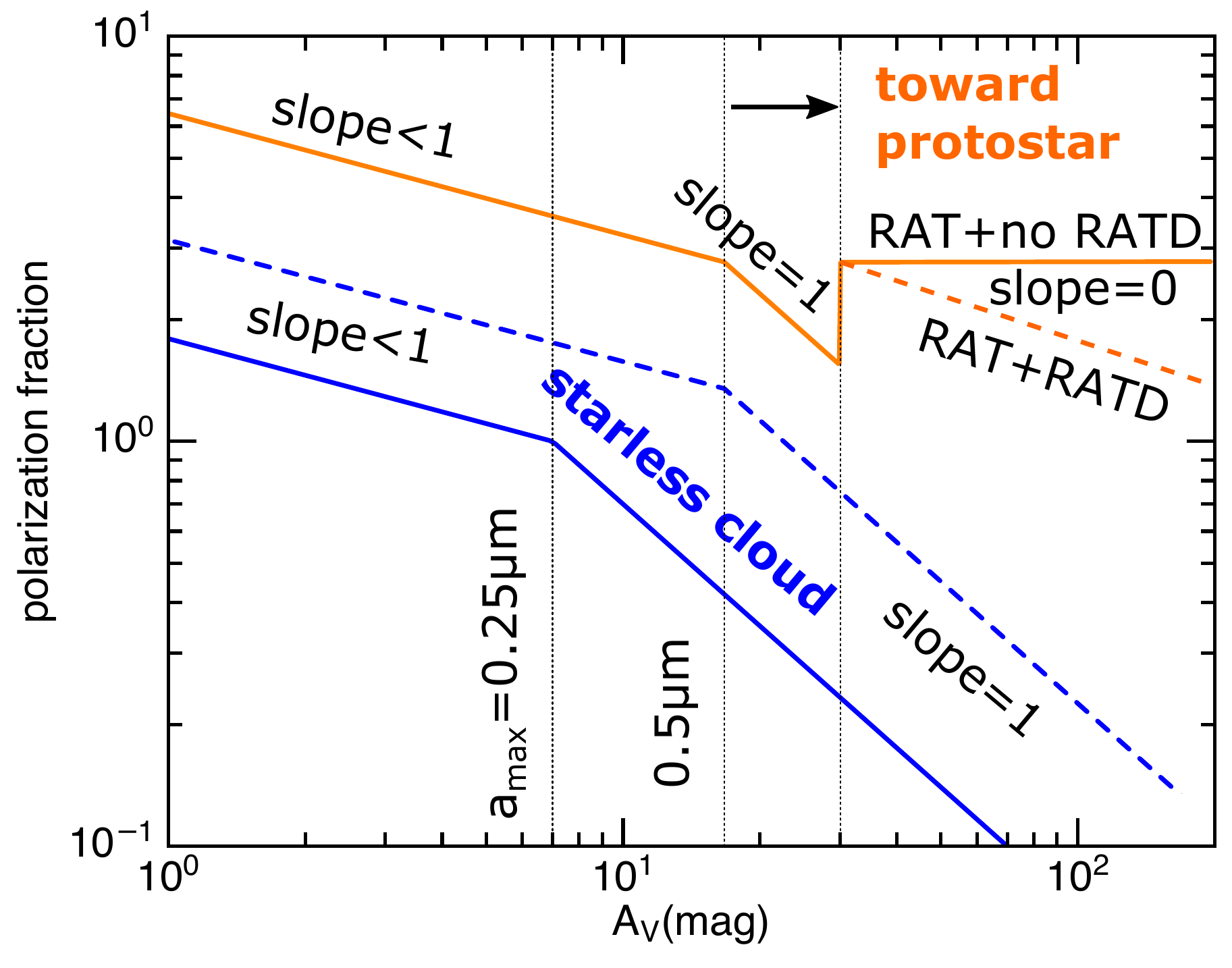}
\caption{Schematic illustration of the variation of the polarization fraction (arbitrary unit) with the visual extinction expected from grain alignment and disruption by RATs, described by $p\propto A_{V}^{-\zeta}$, for a starless core with two values of the maximum grain size (blue lines) and a cloud with an embedded source (orange lines). The slope changes to $\zeta=1$ when grain alignment is completely lost (dotted vertical line). Internal radiation induces alignment of grains near the source, changing the slope from $\zeta=1$ to $\zeta=0$ in the absence of RATD and to $\zeta>0$ in the presence of RATD. The uniform magnetic field geometry is assumed.}
\label{fig:p_AV}
\end{figure}

Figure \ref{fig:p_AV} illustrates the expected variation of the polarization fraction (arbitrary unit) with the visual extinction expected from grain alignment and disruption by RATs for a starless core (blue lines) and a cloud with an embedded protostar (orange lines). In the former case, there exists two slopes, a shallow caused by to the gradual decrease of grain alignment (i.e., increase of $a_{\rm align}$) with $A_{V}$ and a steep one for $A_{V}>A_{V,\max}$ where grain alignment is completely lost. The effect of grain growth shifts the transition point to larger extinction (dashed blue line). In the later case, the protostellar radiation induces alignment of grains near the source and changes the slope from $\zeta=1$ to $\zeta=0$ (solid orange line). In the presence of rotational disruption (RATD), the slope becomes steeper of $\zeta>0$ (dashed orange line). In a realistic situation, there may exist an intermediate period between the steep slope of $\zeta=1$ and $\zeta=0$ where the effect of internal source compensates for the alignment loss by the aISRF.

Finally, note that the average dust temperature $T_{d}$ is usually inferred from fitting SED of thermal dust emission. Therefore, one can take use of this observational parameter to constrain the maximum grain size. Indeed, the radiation strength is approximately given by $U=(T_{d}/T_{d,0})^{6}$, and Equation (\ref{eq:aalign_ana}) yields
\bea
a_{\max}> a_{\rm align} &\simeq&  0.055\hat{\rho}^{-1/7} \left(\frac{n_{3}T_{1}}{\gamma_{-1}}\right)^{2/7}\left(\frac{\bar{\lambda}}{1.2\mum}\right)^{4/7}\nonumber\\
 &&\times\left(\frac{T_{d}}{16.4\K}\right)^{-12/7} \mum,\label{eq:aalign_grow}
\ena
where the infrared damping is omitted due to its subdominance in dense clouds. For example, if observations toward a starless core suggest $T_{d}\sim 15\K$, Equation (\ref{eq:aalign_grow}) implies $a_{\max}>a_{\rm align}=0.45\mum$ and $0.85\mum$ for $n_{\H}=10^{6}, 10^{7}\cm^{-3}$, respectively, assuming $\gamma=0.3$ and $\bar{\lambda}\sim 2\mum$ for the attenuated ISRF. Therefore, one concludes that grain growth must occur in these dense regions if the polarization slope is $\zeta<1$.

\section{Summary}\label{sec:summary}
We study alignment and rotational disruption of dust grains by radiative torques and its implication for understanding the origins of polarization holes in dense MCs. The main findings of our results are summarized as follows:

\begin{enumerate}

\item
Using RAT alignment theory, we derive analytical formulae for the minimum size of aligned grains by RATs as a function of the local parameters, including gas density, temperature, visual extinction $A_{V}$, and the radiation spectrum at the surface of a cloud. Our simple analytical formulae can be used to estimate minimum grain size for alignment and disruption by RATs for arbitrary local physical parameters.

\item We apply our analytical formulae for a dense molecular cloud without embedded sources and find excellent agreement with numerical results. The grain alignment size increases with increasing the local density and the visual extinction from the surface. 

\item We derive an analytical formula for the maximum visual extinction where grain alignment still exists in starless MCs, which depends on the maximum grain size, gas density, and the ambient radiation field.

\item The loss of grain alignment in starless cores is expected to result in the decrease of the polarization with increasing optical depth, which reproduces the polarization hole in starless cores. Therefore, the detection of polarization from dense regions of high extinction implies grain growth to larger than the alignment size.

\item For dense clouds with an embedded source, we demonstrate that the minimum of aligned grains first decreases slowly in the envelope and then accelerates when entering the central core with the constant density. Therefore, the polarization degree of dust emission increases with increasing the emission intensity.

\item We find that rotational disruption of large grains into smaller ones by RATs (i.e., RATD effect) can occur toward the central source. The effect is more efficient for hot cores around protostars.

\item We find that the popular polarization hole observed toward protostars is inconsistent with the classical RAT theory. However, the decrease of polarization induced by the RATD effect can reproduce such a polarization hole without appealing to the magnetic field tangling.

\end{enumerate}

\acknowledgments
We thank C. Hull for interesting discussions on polarization holes. T.H. acknowledges the support by the National Research Foundation of Korea (NRF) grants funded by the Korea government (MSIT) through the Mid-career Research Program (2019R1A2C1087045). L.N.T's research was conducted in part at the SOFIA Science Center, which is operated by the Universities Space Research Association under contract NNA17BF53C with the National Aeronautics and Space Administration. P.N.D. and N.B.N. are funded by Vietnam National Foundation for Science and Technology Development (NAFOSTED) under grant number 103.99-2019.368.


\bibliography{ms.bbl}

\end{document}